\documentclass[fleqn,usenatbib]{mnras}

\usepackage{newtxtext}
\usepackage{mathptmx}

\usepackage{graphicx}
\usepackage{hyperref}
\usepackage{multirow}
\usepackage{booktabs}
\usepackage{siunitx, array}
\usepackage{amsmath}
\usepackage{xcolor}
\usepackage{verbatim}
\usepackage{hyperref}

\usepackage{verbdef}
\usepackage{fancyvrb}
\usepackage{adjustbox}
\usepackage{xcolor}
\usepackage{diagbox}
\usepackage{float,lscape}
\usepackage{newpxmath} 

\newbox\verbbox
\usepackage[flushleft]{threeparttable}

\usepackage[T1]{fontenc}

\DeclareRobustCommand{\VAN}[3]{#2}
\let\VANthebibliography\thebibliography
\def\thebibliography{\DeclareRobustCommand{\VAN}[3]{##3}\VANthebibliography}

\usepackage{graphicx}	

\title[Water ice detection in debris discs]{The characterisation of water ice in debris discs: implications for JWST scattered light observations}

\author[M. Kim et al.]{
Minjae Kim$^{\,1, 2}$\thanks{E-mail: minjae.k.kim@warwick.ac.uk}, Grant M. Kennedy$^{\,1, 2}$, and Veronica Roccatagliata$^{\,3, 4}$\\
$^{1}$Department of Physics, University of Warwick, Gibbet Hill Road, Coventry CV4 7AL, UK \\
$^{2}$Centre for Exoplanets and Habitability, University of Warwick, Gibbet Hill Road, Coventry CV4 7AL, UK\\
$^{3}$Alma Mater Studiorum, Universit$\grave{a}$ di Bologna, Dipartimento di Fisica e Astronomia, Via Gobetti 93/2, 40129 Bologna, Italy\\
$^{4}$INAF-Osservatorio Astrofisico di Arcetri, Largo E. Fermi 5, 50125 Firenze, Italy}

\date{Accepted 2024 August 6. Received 2024 July 12; in original form 2024 April 23}

\pubyear{2024}

\begin{document}
\label{firstpage}
\pagerange{\pageref{firstpage}--\pageref{lastpage}}
\maketitle

\begin{abstract}
Water ice plays a crucial role throughout the different stages of planetary evolution and is abundant in the Universe. However, its presence and nature in debris discs of exoplanetary systems are not yet strongly established observationally. In this study, we quantify and discuss the impact of ice parameters such as volume fraction ${\mathcal{F}}_{\rm ice}$, blow-out grain size, size distribution, and its phase on the observational appearance of debris discs, considering the diverse nature of these systems around stellar spectral types ranging from A to M. Our findings reveal that the prominent ice features at approximately 2.7 and 3.3\,$\mu$m depend on both the water ice fraction ${\mathcal{F}}_{\rm ice}$ and the scattering angle, with backscattering geometries yielding the most prominent signatures. When the phase function is considered and data are not background limited, strong forward and backward scattering (near edge-on discs) are expected to yield the strongest detections in images/spectra for A or F-type stars, while scattering angle matters less for later type stars. The Fresnel peak at 3.1\,$\mu$m serves as a viable discriminant for the transitional phase (crystalline/amorphous), while simultaneously constraining the water ice temperature. For JWST imaging, we find that the F356W and F444W filter combination is most effective for constraining the grain size distribution, while the F356W and F277W filter combination provides better constraints on the ice fraction ${\mathcal{F}}_{\rm ice}$ in debris discs. However, degeneracy between the grain size distribution and ice fraction when using photometric flux ratios means that obtaining robust constraints will likely require more than two filters, or spectroscopic data.
\end{abstract}



\section{Introduction}\label{sec: introduction}

\noindent Water ice (H$_{2}$O, a solid-state form of water; hereafter referred to as ``ice'') bears significant implications for various stages of planetary evolution such as enhancing grain sticking through lower fragmentation velocity and compositional gradients including setting up of ice snow lines, (e.g.,~\citealp{Blum2008, Min2011}). Furthermore, it is fundamental in the maintenance of life as we know it. For example, ice is an important ingredient in molecules as well as planetary atmospheres and surfaces that enable life to form and flourish. While ice is unsurprisingly ubiquitous in the Universe, found on planets, within stellar systems, and even in the vastness of interstellar space as well as on Earth, the origin of water on Earth has remained a long-standing enigma. There are two probable scenarios (e.g., \citealp{vanDishoeck2014, Lunine2003, Leeuw2010, Drake2005, Ikoma2006}): the ``dry scenario'' posits that planets accreted from water-depleted materials within the snowline, followed by water delivery via water-rich comets and asteroids; the ``wet scenario'' proposes that planets accreted a water-rich atmosphere, formed from local planetesimals that retained some ice at high temperatures through physisorption/chemisorption onto silicate grains, or formed beyond the snowline and potentially later migrated inwards (or a time-variable location of the snowline). In particular, Earth's geochemical similarity to comets and meteorites supports the dry scenario (e.g.,~\citealp{Hartogh2011, Lis2019}). Icy planetesimals such as comets likely delivered water molecules and crucial light elements (e.g.,~C, H, and O) to the inner Solar System after planetary cooling (e.g.,~\citealp{Roberts2000, Hartogh2011}). These findings highlight the significance of celestial icy bodies such as asteroids, Kuiper Belt Objects (KBO), and long-period comets as potential reservoirs of life-sustaining water and volatiles. These bodies reside in circumstellar discs around other main sequence stars, known as debris discs, which are analogous to the asteroid and Kuiper Belts in the Solar System, and commonly observed in exoplanetary systems (\citealp{Wyatt2008}).\newline 
\indent While the direct detection of individual small icy bodies in debris disc systems is challenging, observing smaller ($\sim$$\mu$m sized) fragments, which originate from planetesimals during collisions, within these systems can offer valuable insight regarding the composition of parent bodies. However, the presence of ice in debris discs is not strongly established observationally due to the inherently faint and cold (typically $\ll$~100~K) dust with the limitation of current instrumentation in characterising debris disc composition. Only the tentative detection of a broad emission peak around 62~$\mu\rm{m}$ crystalline ice feature has so far been inferred in debris discs around HD 181327 (\citealp{Chen2008}), though the ice was also inferred from mid to far-infrared broadband photometry (e.g., \citealp{Lebreton2012}) and scattered light from the discs (e.g., \citealp{Kueny2024}). \newline
\indent The advent of the James Webb Space Telescope (JWST; \citealp{Gardner2006}) and Extremely Large Telescope (ELT; \citealp{Padovani2023}) presents a transformative opportunity, showing resolved diverse disc structures including exo-asteroid and exo-Kuiper belts (e.g., \citealp{Gaspar2023, boccaletti2023, Rebollido2024}). Moreover, its capabilities potentially extend to observing the prominent near-to-mid infrared (IR) signatures of scattered light, revealing the presence of ice or icy dust (\citealp{Kim2019})\footnote{JWST GO Program 1563 (Icy Kuiper Belts in Exoplanetary Systems, PI: Christine Chen) plans to observe near-infrared reflectance spectra from dust in the Kuiper Belt regions of four nearby debris discs.}. In particular, JWST NIRSpec recently spotted the first clear signs of crystalline water ice, a neatly ordered structure, in the KBO Chariklo in our Solar System (\citealp{Santos-Sanz202}\footnote{JWST GTO Program 1271 (ToO TNOs: `Unveiling the Kuiper Belt by Stellar Occultations', PI: Pablo Santos-Sanz)}). This detection potentially hints at past warming events (\citealp{Jenniskens, Prialnik2022}) as observed in comets undergoing partial crystallisation at perihelion (\citealp{Bar-Nun1985, Moore1992_Hudson1992}), although most of our solar system's comets were formed at temperatures below 50\,K and still consist mostly of irregularly structured amorphous ice (\citealp{Patashnick1974,vanDishoeck2013}). Hence, the presence of crystalline ice suggests micro-collisions, leading to resurfacing events that either expose pristine material or trigger crystallisation processes depending on temperature (\citealp{Jewitt2004}). Furthermore, unlike its amorphous counterpart, crystalline ice is not susceptible to reverting to its disordered state solely due to temperature drops. Instead, it likely requires exposure to constant high-energy particles induced by particle radiation with the solar wind and cosmic rays and photons like UV and X-rays (\citealp{Hansen2004}) to transit its phase again into amorphous ice. On the other hand, the presence of amorphous ice holds additional significance with its unique properties, characterised by lower thermal conductivities and viscosities compared to crystalline ice (\citealp{Jenniskens1998}), resulting in an effective trap and release mechanism for volatiles and other guest molecules. Consequently, variations in ice abundance and phase (amorphous vs. crystalline) serve as a powerful diagnostic tool for probing the thermal history and environmental conditions within debris discs. This, in turn, offers invaluable insights into planet formation and potentially even life itself, through mechanisms like volatile delivery via cometary activity. \newline

\noindent The goal of this study is to answer the key question of how observations may constrain the ice fraction in grains, and the different forms of ice (amorphous and crystalline ice). For this purpose, we conduct a comprehensive numerical feasibility study, assuming a range of ice fractions within icy dust mixtures across various debris disc structures and stellar spectral types. We particularly focus on the shorter wavelength regime (e.g.,~$\sim$1 to 8\,$\mu$m), where scattered light by dust grains dominates. The detection of ice through the scattering of stellar radiation is predominantly viable in systems with lower optical depths, such as debris discs (\citealp{Kim2019}). Furthermore, the disk colour in scattered light reveals the degree of forward scattering, whose strength depends on particle size, and is therefore directly informative about dust grain properties, including their composition and size, and their size distribution (e.g., \citealp{Hughes2018}). In particular, these wavelength regimes, which are of particular interest to JWST (\citealt{Gardner2006}) and ELT (\citealp{Padovani2023}), harbour a wealth of diagnostic features indicative of ice, providing valuable insights and observational constraints on its composition and distribution. This paper is organised as follows: We first discuss the characteristics of two different ice states (i.e.,~amorphous and crystalline ice) in Sect.~\ref{method: determination of ice state}. Next, we describe our model in Sect.~\ref{sec: model}. We present and analyse our results, focusing on the observational characteristics of debris discs including detailed assessments of spectral energy distributions (SEDs) and scattering phase functions (SPFs) in Sect.~\ref{sec: results-SED} and Sect.~\ref{sec: results-Phase functions}. We also discuss the detectability and characterisation of ice in debris discs using JWST spectra and photometry in Sect.~\ref{sec: ice detection with JWST}. We summarise our findings in Sect.~\ref{sec: summary and outlooks}. 

\begin{figure*}
\includegraphics[width=17.8cm, height= 7cm]{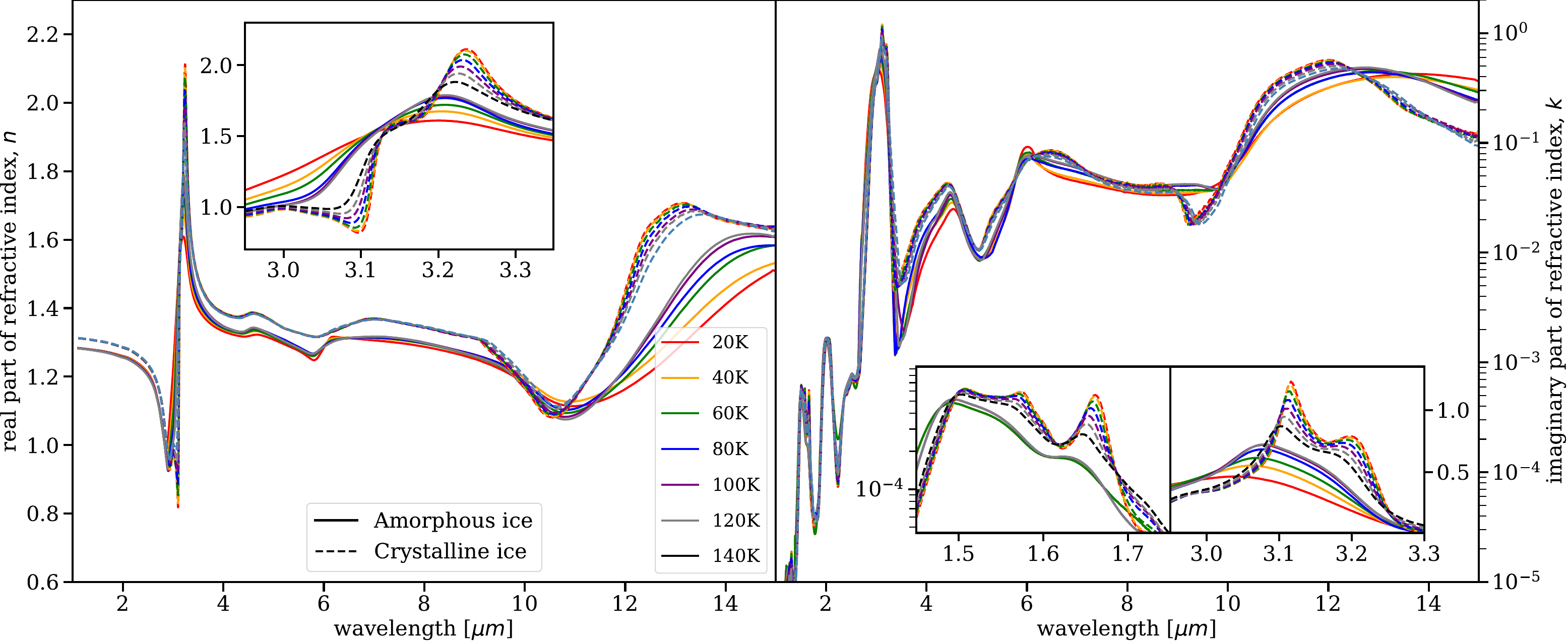}
\caption{Real parts of refractive index $n$ (left) and imaginary parts of refractive index $k$ (right) depending on temperature (\citealp{Mastrapa2009}). \textit{Left}: The real part of the refractive index of crystalline ice shows a large temperature-dependent peak at 3.1 microns and a sharp decrease to a lower peak at 3.25 microns, whereas one of amorphous ice contains a less structured peak which is more continuous throughout that region. \textit{Right}: The depth of the 1.65\,$\mu$m band of the imaginary part of crystalline ice is also a good indicator of crystallinity as it is stronger in colder crystalline ice and almost nonexistent in amorphous ice. See Sect.~\ref{sec: methods} for details.}
\label{fig: n_k}
\end{figure*}

\section{Methods}\label{sec: methods}

\subsection{Optical properties of ice phases and its determination}\label{method: determination of ice state}

In general, there are two different geometric arrangements of molecules in ice: amorphous and crystalline. Amorphous ice is a metastable form with no large-scale regularity in molecular orientations or positions. It can be produced when ice is deposited from vapour at very low temperatures ($\ll$100\,K; \citealp{Klinger1985}). Upon gradual warming to around 130-140\,K (\citealt{Jenniskens, Prialnik2022}), the molecules rearrange into lower energy orientations through an annealing process, releasing energy exothermically and forming fully ordered cubic or hexagonal crystalline ice with lattice structures. 

To identify the two different structural types, several characteristics of strong ice absorption bands are typically investigated, including existence, position, shape, width, and intensity. Fig.~\ref{fig: n_k} describes the real parts of refractive index $n$ (left) and imaginary parts of refractive index $k$ (right) of the complex refractive index $\tilde{n}$ = $n$ + $ik$, depending on ice phase and temperature (\citealp{Mastrapa2009}). Here, $n$ represents the ratio of the speed of light in vacuum and the phase velocity of the light in the medium, which is responsible for scattering, while $k$ is the extinction coefficient, responsible for the emission. In general, wide and relatively unstructured amorphous ice bands shift to shorter wavelengths upon crystallization, becoming narrower, more intense, and often revealing new structures. Furthermore, a weak, smooth signature indicates cold amorphous or warm crystalline ice, whereas a strong, multi-peaked band generally signals cold crystalline ice. 

Of particular interest is the shape of the spectrum in the near-infrared region. In particular, the spectral features ranging from $\sim$3.0 to 3.3\,$\mu$m due to the O-H symmetric and anti-symmetric stretching modes are crucial for distinguishing between the two distinct phases. For example, $n$ and $k$ values of crystalline ice show a large peak at $\sim$3.1\,$\mu$m and a sharp decrease to a lower peak at $\sim$3.2\,$\mu$m, whereas amorphous ice contains a less-structured peak, which is more continuous throughout that region (see Fig.~\ref{fig: n_k}; \citealp{Mastrapa2009}). Furthermore, $n$ and $k$ display temperature-dependent characteristics at $\sim$ 3\,$\mu$m for both amorphous and crystalline ice, albeit with differing behaviours. In general, higher temperatures correspond to lower $n$ and $k$ values for crystalline ice, while the opposite holds for amorphous ice (e.g.,~see inset plots of Fig.~\ref{fig: n_k}). Specifically, the $n$ value remains almost constant for crystalline ice at $\sim$3.0\,$\mu$m, while it exhibits a lower value for amorphous ice at higher temperatures (e.g.,~see the inset plot of the left panel of Fig.~\ref{fig: n_k}). At $\sim$3.1\,$\mu$m, the situation is reversed (i.e.,~$n$ value is relatively constant for amorphous ice but shows lower values for crystalline ice at lower temperatures). For the $k$ values, crystalline ice shows a strong feature compared to amorphous ice and exhibits temperature dependence, that is, higher and narrow peak strength for lower temperatures for crystalline ice, whereas a border one for higher temperatures for amorphous ice (e.g.,~see inset plots of the right panel of Fig.~\ref{fig: n_k}).

The depth of the 1.65\,$\mu$m band of the imaginary part of crystalline ice is also a good indicator of crystallinity as it is deeper in colder crystalline ice and nearly absent in amorphous ice (\citealp{Fink1975, Schmitt1998}). However, this band would not be very useful as the property that distinguishes the type of ice is the depth of the band, not the shape (\citealp{Newman2008}) as amorphous ice also shows weak features that cannot be separated from higher temperature crystalline ice (e.g.,~see the left inset plot of the right panel of Fig.~\ref{fig: n_k}). Furthermore, this characteristic would disappear and all of the spectra would look more or less the same in the region of this band upon normalisation, indicating that this region might contain misleading data (\citealp{Newman2008}). There are also weak absorption features of amorphous ice at around 4.53\,$\mu$m due to the combination mode and 6.06\,$\mu$m due to the overtone of the libration mode (\citealp{Palumbo2005}). 

\begin{table*}
\centering
\def\arraystretch{1.4}             
\caption{Stellar parameters along with corresponding blow-out grain sizes $a_{\rm bo}$ for the simulations and their references, and their corresponding ``ice survival line'' in au and seconds of arc (i.e.,~temperature $\sim$110\,K; so-called ice sublimation radius) as a function of blow-out grain size $a_{\rm \,bo}$ depending on ice fraction $\mathcal{F}_{\rm{ice}}$ (0.1 to 0.9 with a width of 0.1). These calculations assume nearby stellar systems, such as Fomalhaut, located at a distance of 7.7 parsecs (\citealp{Mamajek2012}). The chosen stellar parameters are selected as one of the representative stars which are listed in brackets for each spectral type. See Sect.~\ref{sec: model} for details.}
\renewcommand{\arraystretch}{1.4}
\begin{threeparttable}
\normalsize
\begin{tabular}{llllcc}
\hline\hline
SpT (exemplary system) & Temperature & Radius &  Blow-out size & \multicolumn{2}{c}{Ice survival lines [au] (["])} \\
 & [K] & [R$_\odot$] & $a_{\rm\,bo}$ [$\mu$m] & $a_{\rm{\,bo}}$ & 1000\,$\mu$m \\ \hline
A (Fomalhaut) 		& 8590\tnote{\rm{\,a}}  	& 1.842\tnote{\rm{\,a}}  & 2.37 & 36.79 $\pm$ 1.46 (4.78 $\pm$ 0.19) & 27.09 $\pm$ 0.6 (3.52 $\pm$ 0.08)\\
F (q$^{1}$ Eridani) 	& 6218\tnote{\rm{\,b}}  	& 1.1\tnote{\rm{\,b}} & 0.71 & 14.89 $\pm$ 1.85 (1.93 $\pm$ 0.24) & 8.49 $\pm$ 0.19 (1.10 $\pm$ 0.02)\\
G (HD 207129) 	& 5850\tnote{\rm{\,c}}  	& 0.993\tnote{\rm{\,c}} & 0.45 & 13.72 $\pm$ 2.62 (1.78 $\pm$ 0.34) & 6.79$\pm$ 0.15 (0.88 $\pm$ 0.02)\\
K (Epsilon Eridani) 	& 5116\tnote{\rm{\,d}}  	& 0.74\tnote{\rm{\,d}} & (0.264) & 7.37 $\pm$ 1.54 (0.96 $\pm$ 0.2) & 3.87 $\pm$ 0.08 (0.5 $\pm$ 0.01)\\
M (AU Microscopii)  	& 3665\tnote{\rm{\,e}}  	& 0.82\tnote{\rm{\,e}} & (0.264) & 3.33 $\pm$ 0.58 (0.43 $\pm$ 0.08) & 2.05 $\pm$ 0.04 (0.27 $\pm$ 0.01)\\
\hline
\end{tabular}
\end{threeparttable}
 \begin{tablenotes}
  \item $^{\rm a}$ \citealp{Mamajek2012}, $^{\rm b}$ \citealp{Marmier2013}, $^{\rm c}$ \citealp{Marmier2013}, $^{\rm d}$ \citealp{Watson11}, $^{\rm e}$ \citealp{Donati2023}
 \end{tablenotes}
\label{table: star, disc and dust}
\end{table*}

\subsection{Model description}\label{sec: model}

In the following sections, we discuss the considered model parameters in the present study, which are motivated by observations of currently known debris disc systems. 

While target-specific and instrumental considerations may influence observations of individual sources, we do not consider those specifics in this work, adopting a generalised approach. Our study is based on the assumption that targets selected for ice-related science cases will have sufficiently well-resolved discs that technical considerations do not pose a significant challenge, and that disc spectra or photometry can be extracted and compared with our models. Our findings and conclusions may indeed provide guidance to observers in the design and optimisation of their observational strategies. Given the inner working angle of coronagraphic observations (e.g., JWST NIRCam round/bar occulters, ranging from 0.29 to 0.84 arcseconds at $\sim$ 3 $\mu\rm{m}$; \citealp{Perrin2018}), disks with typical radii of a few tens of au can be resolved at a few tens of parsecs, and larger disks to greater distances.\newline

\noindent\textbf{Central star}: We consider central stars of the following five spectral types, offering a diverse representation of debris disc environments (in brackets exemplary systems are listed): A star (Fomalhaut), F star (q$^{1}$ Eridani), G star (HD 107146), K star (Epsilon Eridani), and M star (AU Microscopii). Table \ref{table: star, disc and dust} summarises the detailed characteristics of stellar parameters along with corresponding blow-out grain sizes $a_{\rm bo}$ (\citealp{Kirchschlager2013}), and ice survival lines as a function of the size of dust grains. Only the most luminous stars have sufficiently distant ice survival lines that imaging might reach regions where ice has sublimated. \newline

\noindent\textbf{Disc geometry and mass}: We consider single and narrow optically thin belts (e.g., \citealp{Booth2023, Roccatagliata2024}). Disc radii are chosen to be large enough that no ice is sublimating, as the sublimation distance for most systems will be too close to the star to be imaged (see Table \ref{table: star, disc and dust}), and interpretation of the results becomes more complicated at smaller radii when e.g. ice has sublimated from small (warmer) but not large (cooler) grains. Thus, our primary focus is on the results of exterior discs~($\leq$~110 K; see Table~\ref{table: star, disc and dust}) to comprehensively characterise the ice features within these systems. To ensure no ice has sublimated, we set all disks to have inner radii of 45\,au (see Table~\ref{table: star, disc and dust}) with a width of 10 au. We examine a range of disc inclinations to investigate how scattering angles influence the prominence of ice features in disc observables. 

We consider a non-infinitely flat disc with a half opening angle of 5\,$^{\circ}$ (\citealp{Kim2019, Bertini2023}). Furthermore, the radial surface density distribution is described by a power-law with an index of $n\,(\rm{r})$ $\propto$ $r^{\rm - 1.5}$ (e.g., \citealp{Krivov2006}).

Lastly, we consider the dust mass of debris discs 10$^{\rm-7}$~${\rm M}_{\,\odot}$ based on previous surveys at sub-mm wavelengths (e.g., typical ranges from $\sim$ 10$^{\rm-9}$ to several 10$^{\rm-7}$\,${\rm M}_{\,\odot}$; \citealp{Greaves2005}). We note that within this model the overall mass of discs does not significantly affect its ice spectral features as long as discs remain optically thin. \newline

\noindent\textbf{Grain size distribution}: Since disc appearance (e.g., colour) in scattered light can inform us about grain sizes \citep{Hughes2018}, we investigate how grain size and its distribution influence disc observables.

We first consider a steady-state collisional cascade, with a power law with $\gamma$ = 3.5 for the differential size distribution d$n\,(a)$ $\propto$ $a^{-\gamma}$ d$a$, where $a$ is the size of the dust grain (\citealp{Dohnanyi}). However, the non-gravitational forces acting on grains (e.g., Poynting-Robertson drag; \citealp{Backman1993}) and collisional evolution may further modify the grain size distribution (e.g.,~\citealp{Thebault2007, Loehne2017, Kim2018, Kim_Wolf_2024}). Thus, the present study takes into account different slopes of the size distribution $\gamma$, ranging from 2.5 to 4.5 in steps of 0.5, to investigate the degeneracy between ice abundance and size distribution.

For the smallest grain size, we specifically adopt a blow-out grain size $a_{\rm\,bo}$ (\citealp{Backman1993}) depending on the spectral type (see Table~\ref{table: star, disc and dust}). For cases without a blow-out size (i.e.,~for central stars' effective temperatures below 5250 K; \citealp{Kirchschlager2013}), we consider a minimum grain size of 0.264\,$\mu$m that has been used in previous studies (e.g.,~\citealp{Loehne2017, Kim2018, Kim_Wolf_2024}; see Table~\ref{table: star, disc and dust}). This minimum grain size aligns with earlier observations of debris discs around M-type stars (e.g.,~\citealp{Matthews2015}) based on the strong stellar wind (\citealp{Plavchan2005}), which exert an influence analogous to radiation pressure forces (\citealp{Augereau2006}). Furthermore, silicate or ice grains comparable to this size are minimally affected by the stellar radiation pressure in debris discs around a solar-type star (\citealp{Cataldi2016}) using the Mie scattering theory (\citealp{Mie1908}). For the largest grain size, we adopt a maximum grain radius of 1\,mm as the contribution of larger grains to the net flux is negligible on disc observables. \newline


\noindent\textbf{Grain chemical composition}: To study the influence of ice parameters on the observational appearance of debris discs, we consider two fundamental types of ice in differing physical states: amorphous and crystalline ice (\citealp{Mastrapa2009}; see Sect.~\ref{sec: methods}) with a bulk density of 1.0 g cm$^{-3}$ (\citealp{Kobayashi2008}). Furthermore, we consider astronomical silicate (referred to as ``astrosil'') with a bulk density of 3.8 g cm$^{-3}$, corresponding to olivine stoichiometry of MgFeSiO$_{4}$ (\citealp{Draine}).

We consider grains with a range of ice fractions, where the ice is mixed with astrosil. To describe the optical properties of this composite material (i.e.,~icy dust), resulting from the optical constants and relative fractions of its components, we employ the effective medium theories (EMT). In particular, we compute the effective refractive index, that is, the scattering and extinction behaviours, using the EMT Bruggeman rule (\citealp{Bruggeman1935}), which is formulated symmetrically with respect to an interchange of materials, making it a more reasonable choice when investigating a wide range of ice fraction. To highlight the differences between the two main EMT methods, we also calculate the EMT Maxwell-Garnett (MG) rule (\citealp{Maxwell-Garnett1906}) for inclusions with different bulk materials (i.e.,~ice inclusion-astrosil matrix particles; \citealp{Kim2019, Stuber_Wolf_2022}). We note that the MG mixing rule is only valid when the volume fraction of inclusions is small (${\mathcal{F}}_{\rm ice}$~$<<$~1) due to its topological assumption. Additionally, there is a concern regarding the treatment of which material is considered the matrix and which is the inclusion (e.g., the "inverse MG mixing rule"; \citealp{Mishchenko2000}) if ice dominates the total material volume. The chemical composition of the icy dust aggregates is defined by the fraction of the total ice volume ${\mathcal{F}}_{\rm ice}$~ ranging from 0 (corresponds to a pure astrosil grain) to 1 (corresponding to pure ice), in steps of 0.1, resulting in bulk densities from 3.8 - 1.0 g cm$^{-3}$ (\citealp{Stuber_Wolf_2022}).\newline

\noindent\textbf{Ice sublimation}: Upon their release from larger parent planetesimals, dust grains experience intense stellar radiation, leading to the depletion of ice and consequent modification of their radial distribution within debris discs, which is supported by several observations of the presence of central clearing in debris discs (\citealp{Jura, Chen2008, Morales2011}) and the change of observed colour (\citealp{Golimowski}). Specifically, ice undergoes immediate sublimation when it reaches temperatures of approximately 100 to 110\,K (\citealp{Kim2019, Hayne2015, Stuber_Wolf_2022}) regardless of grain size (\citealp{Kobayashi2011}). We note that the sublimation temperature does not change notably for the ice-dust mixture (\citealp{Kobayashi2011, Potapov2018b}). Table \ref{table: star, disc and dust} presents the calculated ice survival lines for the debris discs as a function of the blow-out grain size $a_{\rm bo}$ and the largest grain size in our model (i.e., 1 mm), depending on the spectral type of the central stars. 

Since our primary focus is on the disc observables located beyond the ice sublimation radius ($\leq$~110 K; see the model parameter of disc geometry and mass in Sect.\ref{sec: model}) as discussed earlier, we consider the results at a distance of 50 au, regardless of the stellar spectral type (see Table \ref{table: star, disc and dust}). We also note that the results beyond the ice sublimation radius within debris discs remain nearly constant.

Furthermore, our model also employs temperature-dependent optical constants of crystalline/amorphous ice (e.g.,~\citealp{Kim2019}) for debris discs harbouring pure ice (i.e.,~${\mathcal{F}}_{\rm ice}$ = 1.0). For this purpose, we use a simple iterative approach, first calculating the radial temperature distribution assuming the optical constants measured at a temperature of 55K. At a given radial location, we then use the optical properties corresponding to these initial temperature calculations. This procedure is repeated until the temperature at each radial location converges, with the convergence criterion set to a temperature width of 10 K.\newline

\noindent\textbf{Simulation of debris disc observables}: Based on wavelength-dependent optical constants (i.e.,~complex refractive index $n$ and $k$, respectively; see Fig.~\ref{fig: n_k}), we calculate the required optical properties (e.g.,~wavelength-dependent scattering and absorption cross sections $C_{\rm sca}$ and $C_{\rm abs}$; see Figs.~\ref{fig: cabs_sca_sil_ice} and \ref{fig: cabs_sca_ice}) with the tool \textit{miex} (\citealp{Wolf2004}) based on Mie scattering theory (\citealp{Mie1908}). 

For the calculation of observational appearances of debris discs (e.g.,~SEDs and SPFs), we use the radiative transfer software {\it DMS} (\citealp{Kim2018, Stuber_Wolf_2022}), which is optimised for optically thin emission of debris discs. 

We compute scattering phase functions from simulated images of debris discs. The image resolution of $\sim$ 0.02 arcseconds per pixel is adopted in the present study, which is sufficiently fine compared to the resolution of the JWST's NIRCam (0.031 arcseconds per pixel for short wavelength channels) and MIRI (0.1 arcseconds per pixel) instruments. The pixel scale is also chosen to be sufficiently small to avoid line-of-sight issues in highly inclined discs, where disc emission at a range of scattering angles might otherwise be mixed in a single pixel.

\section{Results and Discussion}\label{sec: results}

\noindent In the following sections, we discuss the impact of ice parameters outlined in Sect.~\ref{sec: model} on observational characteristics of debris discs, such as the resulting SEDs in Sect.~\ref{sec: results-SED} and SPFs in Sect.~\ref{sec: results-Phase functions}. We consider how JWST observation can constrain the ice fraction via spectra and photometry in Sect.~\ref{sec: ice detection with JWST}.

\subsection{Spectral Energy Distributions}\label{sec: results-SED}

\begin{figure*} 
\includegraphics[width=17.5cm, height= 22cm]{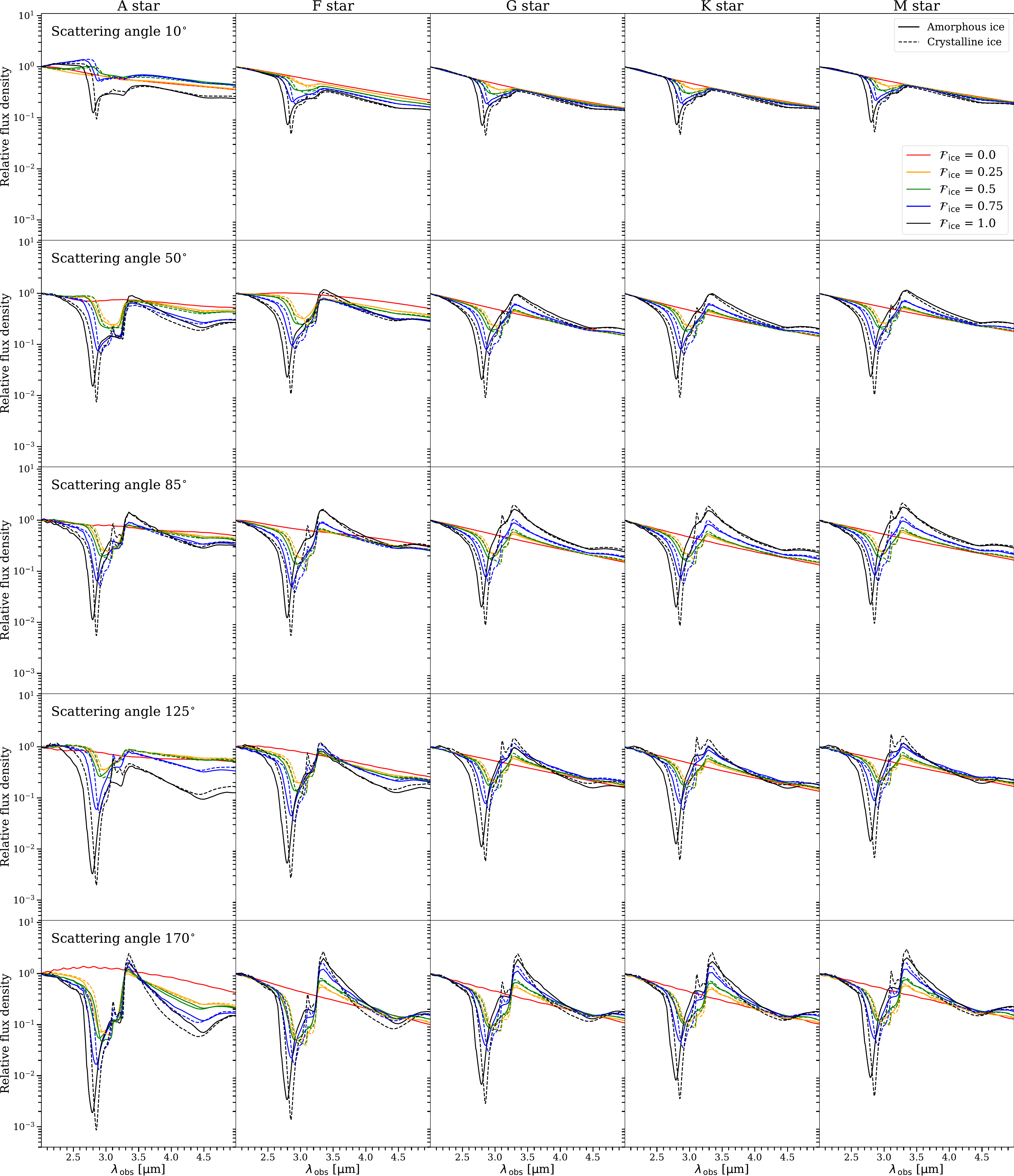}
\caption{Relative flux density of SEDs of exterior debris discs at 50 au (i.e., outside of ice sublimation radius) as a function of the volume fraction of ice ${\mathcal{F}}_{\rm ice}$ and spectral type of the central star depending on scattering angles (i.e., 10, 50, 85, 125, and 170\,$^{\circ}$). ${\mathcal{F}}_{\rm ice}$ = 0 and 1 correspond to pure silicate and pure ice. All spectra are normalised to 1 at 2.0\,$\mu$m. See Sect.~\ref{sec: results-SED} for details.}
\label{fig: SED_AFGKM_40au_RF_sa}
\end{figure*}

\begin{figure*}
\includegraphics[width=18cm, height= 4.5cm]{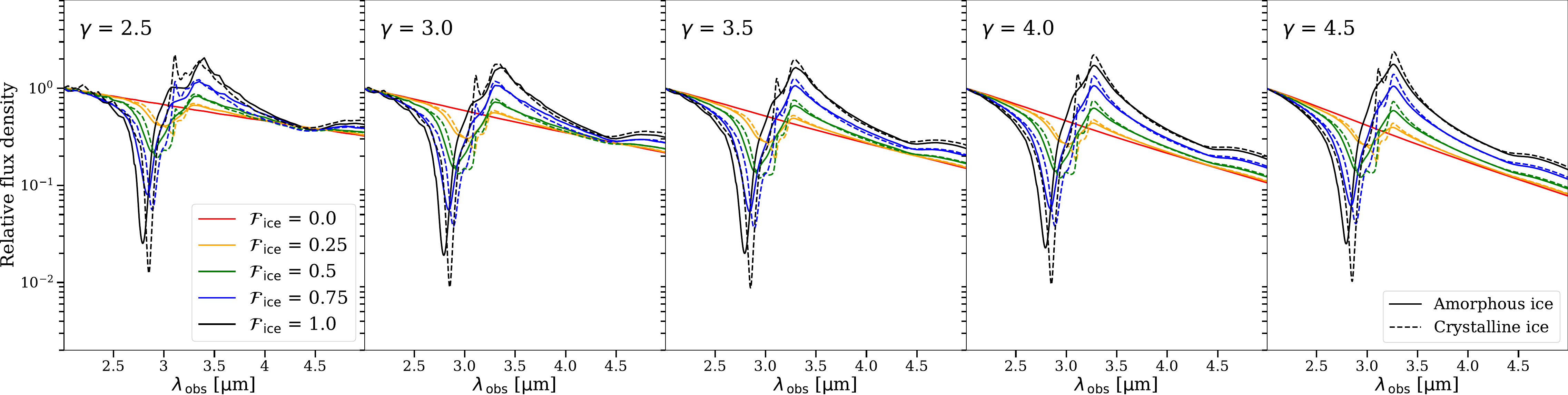}
\caption{Relative flux density of SEDs of exterior debris discs at 50 au around a G star with the scattering angle of 85\,$^{\circ}$ as a function of the volume fraction of ice ${\mathcal{F}}_{\rm ice}$  depending on the power-law index $\gamma$ of the grain size distribution. All spectra are normalised to 1 at 2.0\,$\mu$m. See Sect.~\ref{sec: results-SED} for details.}
\label{fig: SED_different_r_G_2}
\end{figure*}

We investigate the influence of the fractional ratio of ice ${\mathcal{F}}_{\rm {\,ice}}$ and the ice phase in icy dust aggregates, on the resulting SEDs. Our findings are primarily presented with respect to scattering angles -- the angle at which an incident light from the star is deflected by dust particles -- to identify which phase angles are significant for observation, given that integrated disc quantities may not adequately represent data from resolved coronagraphic observations. Fig.~\ref{fig: SED_AFGKM_40au_RF_sa} presents the relative flux density of SEDs of exterior debris discs at 50 au (i.e., outside of ice sublimation radius; see Table~\ref{table: star, disc and dust}) as a function of the volume fraction of ice ${\mathcal{F}}_{\rm ice}$ and spectral type of the central star depending on scattering angles (i.e., 10, 50, 85, 125, and 170\,$^{\circ}$). 

We find that the ice features around $\sim$3\,$\mu$m exhibit a dependence on the scattering angle and thus the disc inclination angle (see Fig.~\ref{fig: SED_AFGKM_40au_RF_sa}). In particular, backscattering (e.g., large scattering angles; shown in the fifth rows of Fig.~\ref{fig: SED_AFGKM_40au_RF_sa}) yields the most prominent ice features. The ice absorption feature, for instance, exhibits an enhancement of approximately two orders of magnitude compared to the forward scattering regime (e.g., small scattering angles; shown in the first rows of Fig.~\ref{fig: SED_AFGKM_40au_RF_sa}). Furthermore, the ice features around 4.5\,$\mu$m also exhibit a scattering angle dependence, which is particularly pronounced for the backscattering. Thus, our findings indicate that forward scattering tends to diminish the prominence of ice features, which aligns with previous studies that have demonstrated a weaker dependence on particle characteristics for forward scattering (e.g., \citealt{Grynko2004}). Consequently, highly inclined discs, which offer a range of substantial backscattering angles, potentially represent optimal scenarios for detecting and characterising the prominent ice features. Backscattering is however overall fainter than forward scattering; we revisit this trade-off below (Sect.~\ref{sec: results-Phase functions}). In the case of spatially resolved observations targeting only the ansae of discs, e.g., observations with a smaller field of view (FOV) such as the JWST NIRSpec IFU (\citealp{Jakobsen2022}), the outcomes are expected to closely resemble those obtained from observations of face-on disc configurations (i.e., scattering angles close to 90$^\circ$), irrespective of the actual disc inclination angle. 

We also find that the contrast within the ice feature (e.g., the strength of the $\sim$ 2.7 and 3.3\,$\mu$m ice features) increases with increasing ${\mathcal{F}}_{\rm {\,ice}}$, particularly pronounced for the backscattering regimes. Additionally, our results indicate that the locations of amorphous and crystalline ice features at $\sim$ 2.7 and 2.8\,$\mu$m shift towards shorter wavelengths as the ice volume fraction ${\mathcal{F}}_{\rm {\,ice}}$ increases. 

Of particular interest is the crystalline ice features at 3.1 (the so-called `Fresnel peak'; \citealp{Brown2006}) and 3.25\,$\mu$m -- the latter being typically weaker. These features serve as viable discriminants for the transitional phase, such as the crystalline ice fraction (crystallinity), while simultaneously offering a temperature probe because this feature is sensitive to temperature (\citealt{Mastrapa2009}), as discussed in Sect.~\ref{sec: methods}. We find that both crystalline ice features require spectral resolution $R$ (= $\lambda$/$\Delta\lambda$) $\sim$ 30 for discrimination. Interestingly, our results suggest that the location of these features, unlike other ice features, remains unchanged across varying levels of ice fraction ${\mathcal{F}}_{\rm ice}$. Furthermore, this feature can persist even with forward scattering. The 3.25\,$\mu$m feature is particularly discernible for a higher fraction of ice in dust from backscattering regimes (e.g., see the blue and black lines in the bottom of Fig.~\ref{fig: SED_AFGKM_40au_RF_sa}), which are pronounced in discs around fainter stars. Consequently, discs harbouring crystalline ice are expected to show at most two distinct features at around $\sim$ 3.1 to 3.4\,$\mu$m wavelength ranges, which is more pronounced in discs around less luminous stars. In contrast, those with amorphous ice typically exhibit a single feature around only $\sim$ 3.4\,$\mu$m. Furthermore, the locations of ice absorption features at $\sim$ 2.8\,$\mu$m, shifting towards shorter wavelengths as well, are further enhanced with a decrease in crystallinity, possibly serving as an additional indicator of the ice composition and distribution within the discs.

The luminosity of intermediate-mass stars (e.g., A-type) effectively expels smaller grains through stellar pressure forces (\citealp{Kirchschlager2013}), resulting in a larger blow-out size (e.g., $a_{\rm\,bo}$ = 2.37\,$\mu$m). However, given that the emission of dust particles with sizes comparable to the observing wavelength is most efficient, we find that this absence of tiny grains, particularly those smaller in size than the ice feature wavelengths (e.g., $\sim$3\,$\mu$m) does not significantly affect the strength or location of the ice features around 3\,$\mu$m. In particular, the prominent ice absorption bands remain intact (see the first column of Fig.~\ref{fig: SED_AFGKM_40au_RF_sa}) although the overall continuum level may change only moderately with a different blow-out size. Thus, we conclude that the blow-out grain size is not a crucial factor in determining the detectability of ice spectral features from SED observations.

Fig.~\ref{fig: SED_different_r_G_2} presents the relative flux density of SEDs for exterior debris discs at a distance of 50 au around a G type star for different power-law indices $\gamma$ of the grain size distribution, at a scattering angle of 85\,$^{\circ}$ (i.e., corresponding to a nearly face-on disc orientation) as a function of the ice volume fraction ${\mathcal{F}}_{\rm ice}$ (see also Sect.~\ref{sec: model}). We find that the grain size distribution directly influences the slope of the SEDs. Higher $\gamma$ values, indicating steeper grain size distributions, result in steeper SED slopes (e.g., \citealp{Kim_Wolf_2024}). However, the strength of most ice features remains relatively consistent across different $\gamma$ values. An exception is the 3.1 $\mu$m crystalline ice feature, which shows sensitivity to $\gamma$. Specifically, shallower grain size distributions (lower $\gamma$ values, e.g., $\gamma$ = 2.5) tend to produce a more prominent 3.1 $\mu$m feature, especially at higher ice fractions ${\mathcal{F}}_{\rm ice}$. In contrast, the 3.3 $\mu$m feature remains relatively unaffected by changes in $\gamma$.


Fig.~\ref{fig: SED_AFGKM_40au_RF_sa_MG} in Appendix~\ref{app: Selection of mixing rules} present relative flux density of SEDs of exterior debris discs at 50 au as a function of the volume fraction of ice ${\mathcal{F}}_{\rm ice}$ and spectral type of the central star depending on scattering angles, using the MG mixing rule (\citealp{Bruggeman1935}; see also Sect.~\ref{sec: model}). We find that the choice of EMT mixing rule, such as the Bruggeman or MG approach, has a relatively minor influence on SEDs (see also Fig.~\ref{fig: SED_AFGKM_40au_RF_sa}, using Bruggeman mixing rule). The primary distinction is observed in the depth of the ice features, where the MG mixing rules tend to produce slightly deeper absorption features. Additionally, the 3.1 $\mu$m crystalline ice feature appears more pronounced and sharper when the MG mixing rules are adopted. Furthermore, the MG mixing rules result in a slight shift of the ice features towards shorter wavelengths, although these effects are hardly discernible. Thus, we conclude that the choice of mixing rule does not dominate or significantly impact the overall spectral features.

\begin{figure*}
\includegraphics[width=18cm, height= 8.5cm]{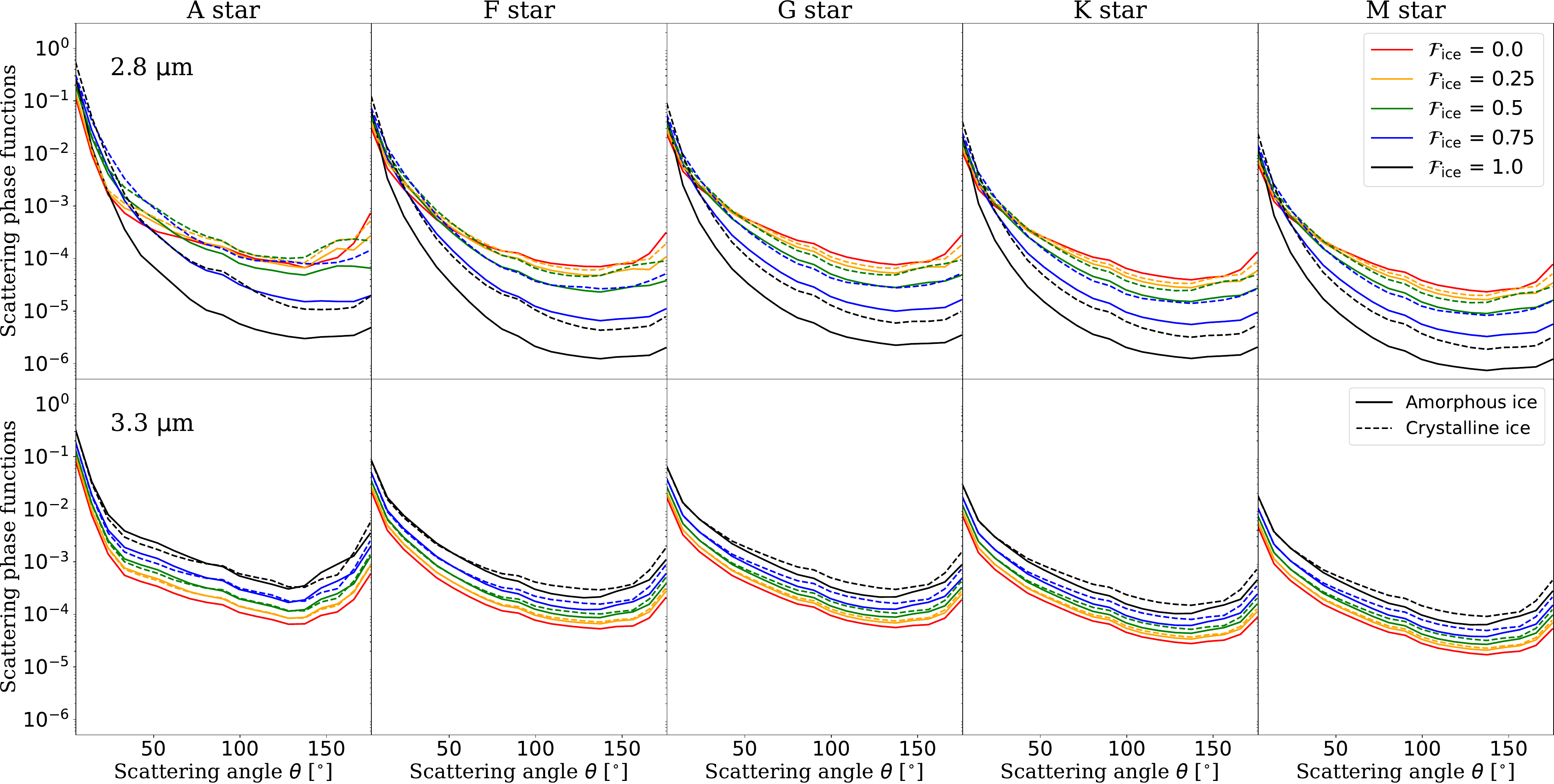}
\caption{Simulated debris disc SPFs as a function of scattering angle and spectral type at 50 au (i.e., outside of ice sublimation radius) harbouring amorphous and crystalline ice-dust aggregate at 2.8 and 3.3\,$\mu$m depending on the ice fraction ${\mathcal{F}}_{\rm ice}$. See Sect.~\ref{sec: results-Phase functions} and Fig.~\ref{fig: SED_AFGKM_40au_RF_sa} for details.}
\label{fig: SPF_AFGKM}
\end{figure*}

\begin{figure*}
\includegraphics[width=18cm, height= 7.8cm]{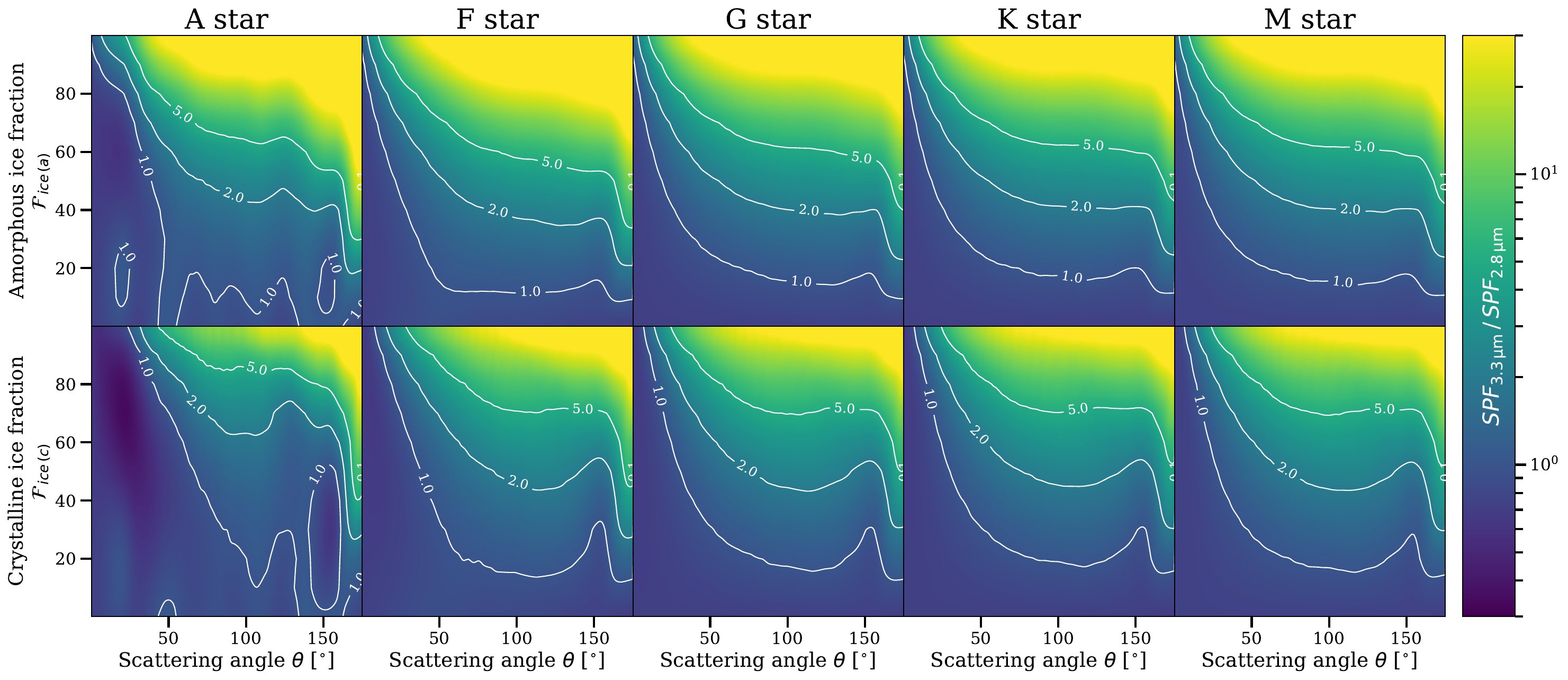}
\caption{The ratio between SPFs of debris discs harbouring amorphous ice-dust aggregate (top panels) and crystalline ice-dust aggregate (bottom panels) at 3.3 and 2.8\,$\mu$m depending on the ice fraction ${\mathcal{F}}_{\rm ice}$ and spectral type. Note that the ratio between SPFs of debris discs harbouring the pure silicate grain (i.e., ${\mathcal{F}}_{\rm ice}$ = 0.0) results in $\sim$ 0.9 (forward scattering angles for A, F, G, K, and M stars) - 1.2 (backscattering for A star). See Sect.~\ref{sec: results-Phase functions} for details.}
\label{fig: SPF_grids}
\end{figure*}

\begin{figure*}
\includegraphics[width=18cm, height= 7.8cm]{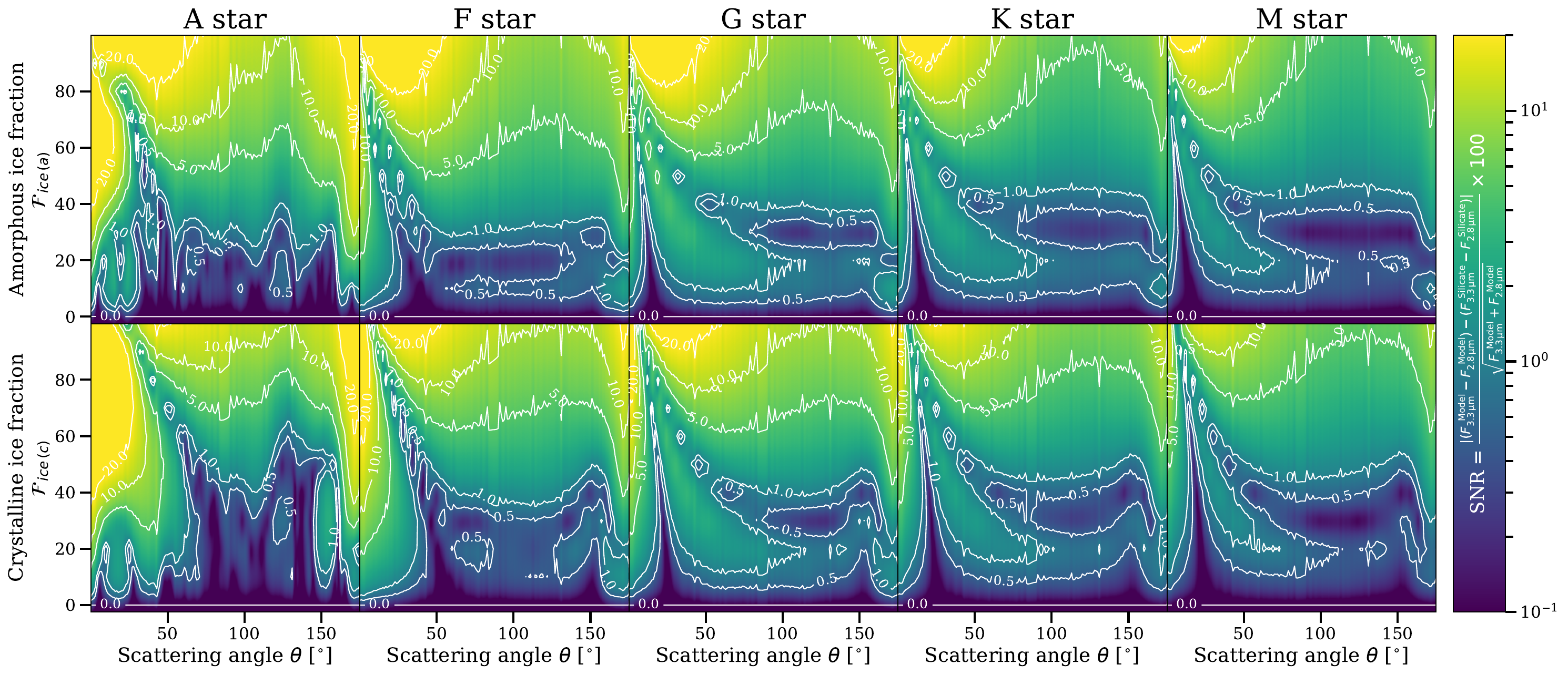}
\caption{The relative SNR (i.e., the signal strength of the ice feature/combined noise from both channels) of debris discs harbouring amorphous ice-dust aggregate (top panels) and crystalline ice-dust aggregate (bottom panels) as a function of the ice fraction ${\mathcal{F}}_{\rm ice}$ depending on the spectral type of central stars. A higher SNR value indicates better resolution and detection capability for the ice features around 3\,$\mu$m. See Sect.~\ref{sec: results-Phase functions} for details.}
\label{fig: spf_ratio_grids_horizontal_SNR}
\end{figure*}



\subsection{Scattering Phase Functions}\label{sec: results-Phase functions}

As discussed in the previous Sect.~\ref{sec: results-SED}, the characteristics of ice features (e.g., strength and width of the peak), particularly those around the $\sim$3\,$\mu$m wavelength region, are influenced not only by the ice parameters such as the ice fraction ${\mathcal{F}}_{\rm ice}$ but also by the given disc inclination (i.e., the corresponding scattering angles). To further investigate this, we present the wavelength-dependent angular distribution of scattered light, focusing on wavelengths corresponding to key ice features. Fig.~\ref{fig: SPF_AFGKM} shows the SPFs as a function of scattering angle for debris discs at 50 au (i.e., outside the ice sublimation radius; see Table~\ref{table: star, disc and dust}) at 2.8 (top panels) and 3.3\,$\mu$m (bottom panels), depending on the ice fraction ${\mathcal{F}}_{\rm ice}$ and the spectral type of the central stars. We note that the SPFs provide the same information as the SEDs (see Fig.~\ref{fig: SED_AFGKM_40au_RF_sa}) but are a different way of visualising the scattering properties of the discs. Given an observation with access to a range of scattering angles, SPFs can provide an additional constraint for the determination of the ice parameters (e.g.,~ice fraction ${\mathcal{F}}_{\rm ice}$), potentially inferring dust compositions, and elucidating the underlying grain size distribution as various characteristics of dust grains (e.g., size and composition) influence the degree of asymmetric light scattering by dust at a given wavelength, altering discs' apparent albedo and thus impacting the observational appearance of debris discs. 

We find that pure silicate (${\mathcal{F}}_{\rm ice}$ = 0.0) in debris discs results in similar scattering functions at 2.8 and 3.3\,$\mu$m. As the ice fraction ${\mathcal{F}}_{\rm ice}$ increases, the phase functions exhibit more prominent features, such as sharper forward-scattering peaks and shallower but lower backscattering behaviour at 2.8\,$\mu$m (e.g., see the top panel of Fig.\ref{fig: SPF_AFGKM}), showing again that the absorption gets stronger for backscattering. Additionally, there is a tendency for the SPF to decrease as amorphous ice content increases in the backscattering regime at 2.8\,$\mu$m, which is not well pronounced at 3.3\,$\mu$m. 

Fig.\ref{fig: SPF_grids} describes the ratio between two SPFs (i.e., SPF$\,\rm_{3.3\,\mu{m}}$ and SPF$\,\rm_{2.8\,\mu{m}}$) of debris discs harbouring amorphous (top panels) and crystalline (bottom panels) ice-dust aggregate as a function of the ice fraction ${\mathcal{F}}_{\rm ice}$ depending on the spectral type of central stars through grid plots, which shows a way to summarise absorption seen in a spectrum to estimate the ice fraction ${\mathcal{F}}_{\rm ice}$. As discussed earlier, the scattering features of lower ice fractions (e.g., pure silicate) almost remain constant at both 2.8 and 3.3\,$\mu$m, showing a similar ratio $\sim$0.9, which increases with an increase of stellar luminosity, scattering angle and ice fraction ${\mathcal{F}}_{\rm ice}$ and phase. In general, the ratios show almost flat behaviour for scattering angles larger than $\sim$60\,$^{\circ}$ and smaller than $\sim$120\,$^{\circ}$. This indicates again backscattering exhibits more pronounced features and a stronger dependence on the ice fraction ${\mathcal{F}}_{\rm ice}$ (see also Sect.~\ref{sec: results-SED}). Thus, forward scattering shows less pronounced features and weaker dependence on ice fraction but is intrinsically brighter than backscattering. Consequently, a trade-off exists, and it might be that the fainter far side of the disk actually provides a less expensive path to quantifying the ice fraction.

To quantify and thus compare the relative strengths between forward and backscattering, we calculate an SNR by taking the absolute value of the difference in signal strength between the ice features and the combined noise from both channels, assuming Poisson noise statistics. This SNR is a relative quantity for the ice feature at a given scattering angle, and not related to how much signal might be in a specific JWST observation. Significant background flux relative to the fainter backscattering side of the disc may result in non-detection, in which case spectral features would rely on a stronger detection on the forward scattering side. The calculation is performed as follows:



\begin{equation}
SNR = \frac{|(F_{\rm{\,3.3\,\mu{m}}}^{\rm\,\,Model} - F_{\rm{\,2.8\,\mu{m}}}^{\rm\,\,Model}) - (F_{\rm{\,3.3\,\mu{m}}}^{\rm\,\,Silicate} - F_{\rm{\,2.8\,\mu{m}}}^{\rm\,\,Silicate})|}{\sqrt{{F_{\rm{\,3.3\,\mu{m}}}^{\rm\,\,Model}}+ {F_{\rm{2.8\,\mu{m}}}^{\rm\,\,Model}}}} \times 100,
\end{equation}

\noindent where $F$ represents the flux at each wavelength for the models and silicate-only (${\mathcal{F}}_{\rm ice}$ = 0) cases, computed at each scattering angle, and the factor of 100 is included to bring the SNR closer to unity. The SNR estimate is relative to a pure silicate model, and uses spectra that are not normalised across scattering angles, so includes the effect that the backscattering side is fainter (but has stronger ice features).
Consequently, a higher SNR value would indicate better detection capability for the ice features around 3\,$\mu\rm{m}$, as long as the disk is detected at all scattering angles.

Fig.~\ref{fig: spf_ratio_grids_horizontal_SNR} shows the simulated grid plots of the relative SNR. We find that very forward scattering and backscattering are almost equally favourable for the observation of discs with high ice fractions (e.g., ${\mathcal{F}}_{\rm ice}$~>~0.7) and low ice fractions (e.g., ${\mathcal{F}}_{\rm ice}$~<~0.3), particularly around more luminous stars, but obtaining such information may be challenging (e.g., \citealp{Engler2019}), and it is unlikely that very high ice fractions will be detected. For more moderately forward scattering angles, the trends are weaker, with a general preference for forward scattering, but similar SNR values for a given ${\mathcal{F}}_{\rm ice}$. For example, for discs with moderate ice fractions (e.g., ${\mathcal{F}}_{\rm ice}$~$\sim$~0.5), moderately forward scattering exhibits higher SNR values, making greater disc inclinations somewhat more favourable for ice characterisation. Consequently, the optimal scattering geometry depends on spectral type, and to some degree the specific ice fraction ${\mathcal{F}}_{\rm ice}$. In general however, the scattering angle vs. feature strength trade-off yields approximately the same results as a function of scattering angle, and the disc geometry does not necessarily need to be a major consideration for detecting and characterising ice in debris discs with a flux ratio between 2.8 and 3.3\,$\mu$m.

\subsection{Ice detection and characterisation via JWST spectra and photometry}\label{sec: ice detection with JWST}

\begin{figure*}
\includegraphics[width=18cm, height= 11cm]{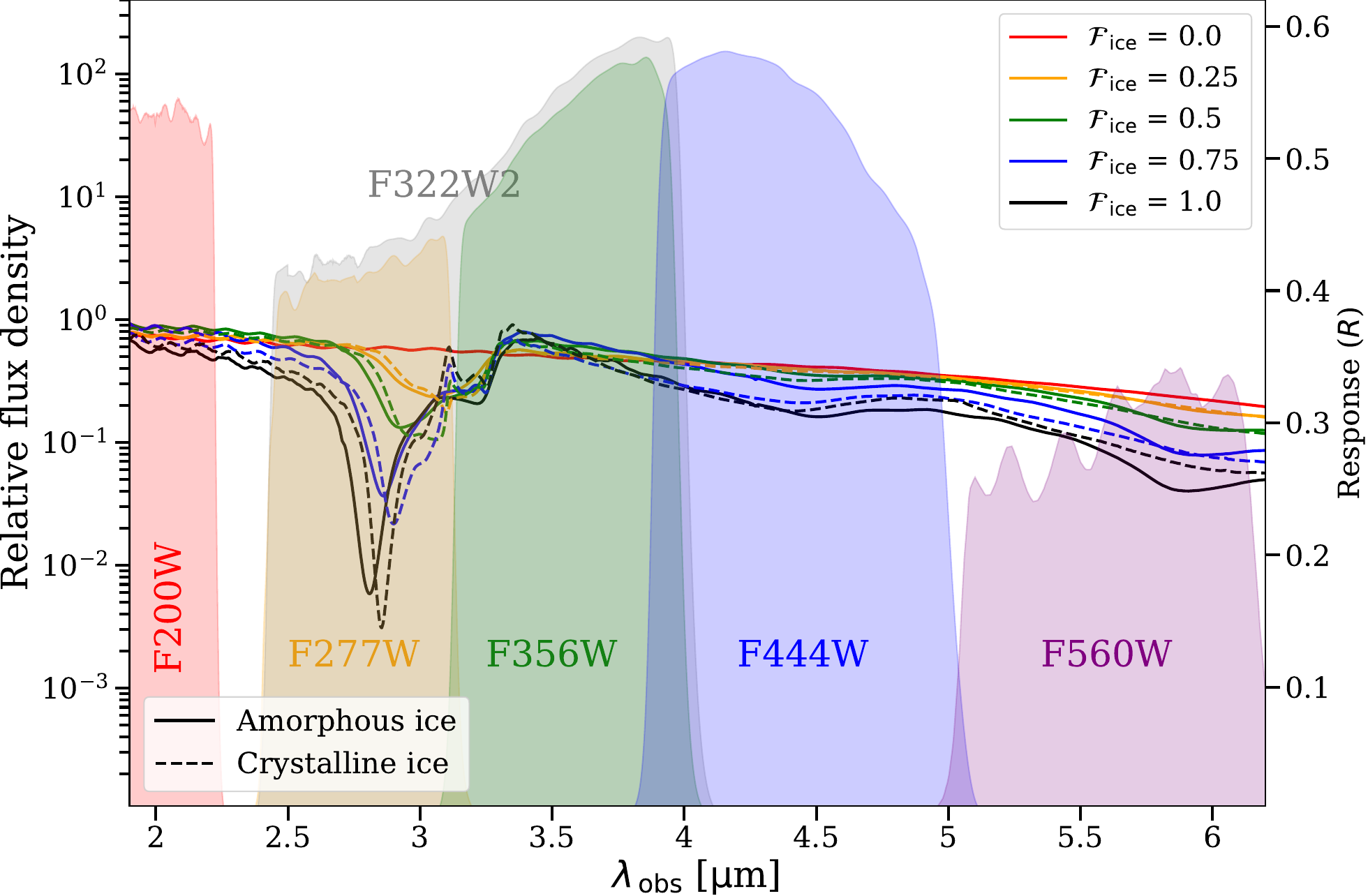}
\caption{Calculated relative flux density of face-on debris disc around A star as the function of ice volume fraction in dust aggregates, incorporating essential NIRCam and MIRI broadband filters to distinguish ice features. See Sect.~\ref{sec: ice detection with JWST} for details.}
\label{fig: SED_filters}
\end{figure*}

To effectively distinguish ice features from observables using JWST imaging filters, we present our results incorporating characteristic lines and JWST NIRCam and MIRI broadband filters. Fig.~\ref{fig: SED_filters} shows the calculated relative flux density of face-on debris discs around an A star as a function of ice volume fraction in dust aggregates, incorporating NIRCam and MIRI broadband
filters. These include the F200W broadband in the short wavelength range (0.6 - 2.3\,$\mu$m), four broadbands (F277W, F322W2, F356W, and F444W) in the long wavelength (2.4 - 5.0\,$\mu$m) channels of NIRCam, and one MIRI broadband filter (F560W) covering the range between 5 and 6.5\,$\mu$m. As shown in Fig.~\ref{fig: SED_filters}, some of these filters are particularly sensitive to the electronic transitions and absorption features associated with ice, potentially providing coarse information about the ice composition and grain properties.

To determine optimal filter combinations, those for which the flux ratio is greatest for a given model, we calculate the flux ratios for all combinations of JWST filters corresponding to ice features, considering all ice models and scattering angles This calculation considers all ice models and scattering angles using the following equation:
\begin{equation}
\frac{\int F  R_{\rm{\,filter}}\,\mathrm{d}\lambda}{\int R_{\rm{\,filter}}\,\mathrm{d}\lambda},  
\end{equation}
where $R_{\rm{\,filter}}$ represents the response function of each JWST filter, respectively. Figures \ref{fig: corner_plot_A_sc}, \ref{fig: corner_plot_F_sc}, \ref{fig: corner_plot_G_sc}, \ref{fig: corner_plot_K_sc}, and \ref{fig: corner_plot_M_sc} in Appendix \ref{app: Photometry with JWST filters} present the calculated corner plots of photometric measurements of debris discs for different spectral types of the central stars. 

The narrow filter centred on ice features (i.e., F277W) is the best, but not the only viable option for constraining ice characterisation. The combination between the F277W and F356W (and/or F322W2) NIRCam filters, and one between the F356W and F444W NIRCam filters, constitute advantageous filter combinations for constraining the ice fraction within the observed dust grains in debris discs (see shaded regions of Figs.~\ref{fig: corner_plot_A_sc}, \ref{fig: corner_plot_F_sc}, \ref{fig: corner_plot_G_sc}, \ref{fig: corner_plot_K_sc}, and \ref{fig: corner_plot_M_sc}).

Fig.~\ref{fig: filter_ice_fraction_AGM_sc} describes the selected filter ratios showing the best performance in constraining the ice fraction, i.e., F356W and F277W (upper panel\footnote{Note that this value is flipped compared to the one (i.e., F277W/F356W) shown in Figs. \ref{fig: corner_plot_A_sc}, \ref{fig: corner_plot_G_sc}, and \ref{fig: corner_plot_M_sc}.}) and F356W and F444W (lower panel), for debris discs around A, G, and M stars. These filter combinations exhibit higher sensitivity to variations in the ice fraction across the entire range of scattering angles considered in our simulations. The F356W/F277W ratio (see the top panels of Fig.~\ref{fig: filter_ice_fraction_AGM_sc}) is potentially the most sensitive to changes in the ice fraction $\mathcal{F}_{\rm ice}$. Furthermore, regarding how rapidly the ratios deviate from the baseline (e.g., pure silicate case, showing almost flat behaviour depending on scattering angles for most of combination) as the ice fraction increases, which is indicative of the minimum detectable ice fraction, it appears that the F277W/F356W ratio is highly sensitive, while the F200W/F277W ratio also performs well in certain scenarios (see the bottom and 5th columns of Figs.~\ref{fig: corner_plot_A_sc}, \ref{fig: corner_plot_F_sc}, \ref{fig: corner_plot_G_sc}, \ref{fig: corner_plot_K_sc}, and \ref{fig: corner_plot_M_sc}). This effect is particularly noticeable for discs around G-type stars and less luminous stars. In general, any combination involving the F277W filter provides the best constraints on the ice fraction $\mathcal{F}_{\rm ice}$, including the minimum detectable ice fraction. Notably, our models suggest that to detect an ice fraction of 10\% in debris discs around G star, utilising the F356W/F277W filter ratio at a 3-sigma significance level, a relative photometric precision of 1.80 - 1.835 \% or better is required.

Additionally, we find that the F356W/F444W ratio (see the bottom panels of Fig.~\ref{fig: filter_ice_fraction_AGM_sc}) also exhibits good sensitivity to the ice fraction $\mathcal{F}_{\rm ice}$, but with a relatively small variation in the flux ratio as $\mathcal{F}_{\rm ice}$ increases from zero. This effect is particularly noticeable for discs around more luminous stars, such as A-type stars. For observations with the F356W/F444W filter, a less stringent photometric precision of 2.297 - 2.433 \% or better is required to achieve the same 3-sigma detection of a 10\% ice fraction in debris discs around G star in our models.

We note that combining observations from the F356W NIRCam filter and the F560W MIRI broadband filters could provide enhanced constraints on the ice fraction, particularly within certain scattering angle regimes (as shown in the first columns of the second rows of figures in Appendix \ref{app: Photometry with JWST filters}). This highlights the potential advantage of leveraging multiple instruments aboard the JWST to facilitate a more comprehensive characterization of the ice fraction present in the observed systems. However, the F560W filter is not designed for coronagraphic observations, which may limit its applicability or suitability for certain observational scenarios.

\begin{figure*}
\includegraphics[width=18cm, height= 7.8cm]{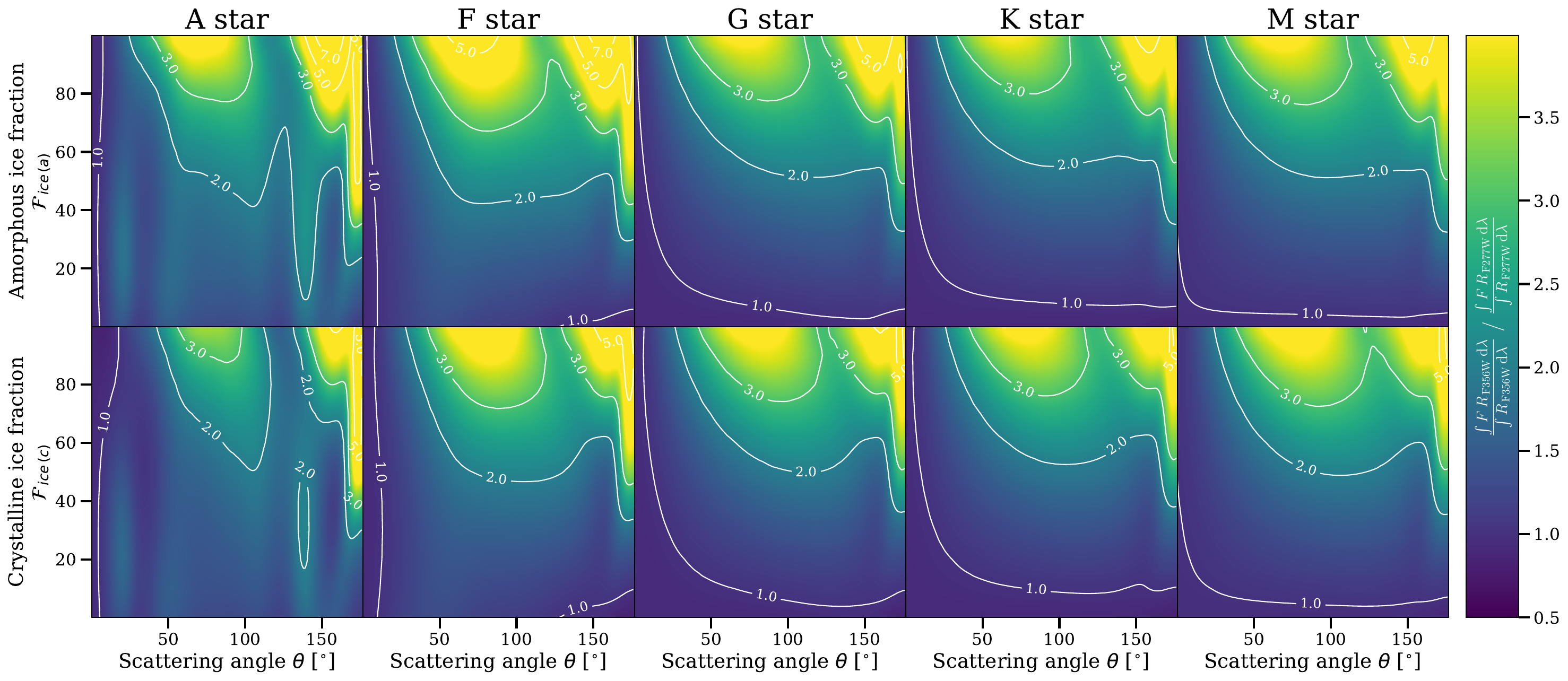}
\caption{The F356W/F277W ratio between SPFs of debris discs harbouring amorphous ice-dust aggregate (top panels) and crystalline ice-dust aggregate (bottom panels), as a function of the ice fraction ${\mathcal{F}}_{\rm ice}$ and spectral type depending on the scattering angle. See Sect.~\ref{sec: results-Phase functions} for details.}
\label{fig: SPF_photometry_grids_356_277}
\end{figure*}

\begin{figure*}
\includegraphics[width=18cm, height= 7.8cm]{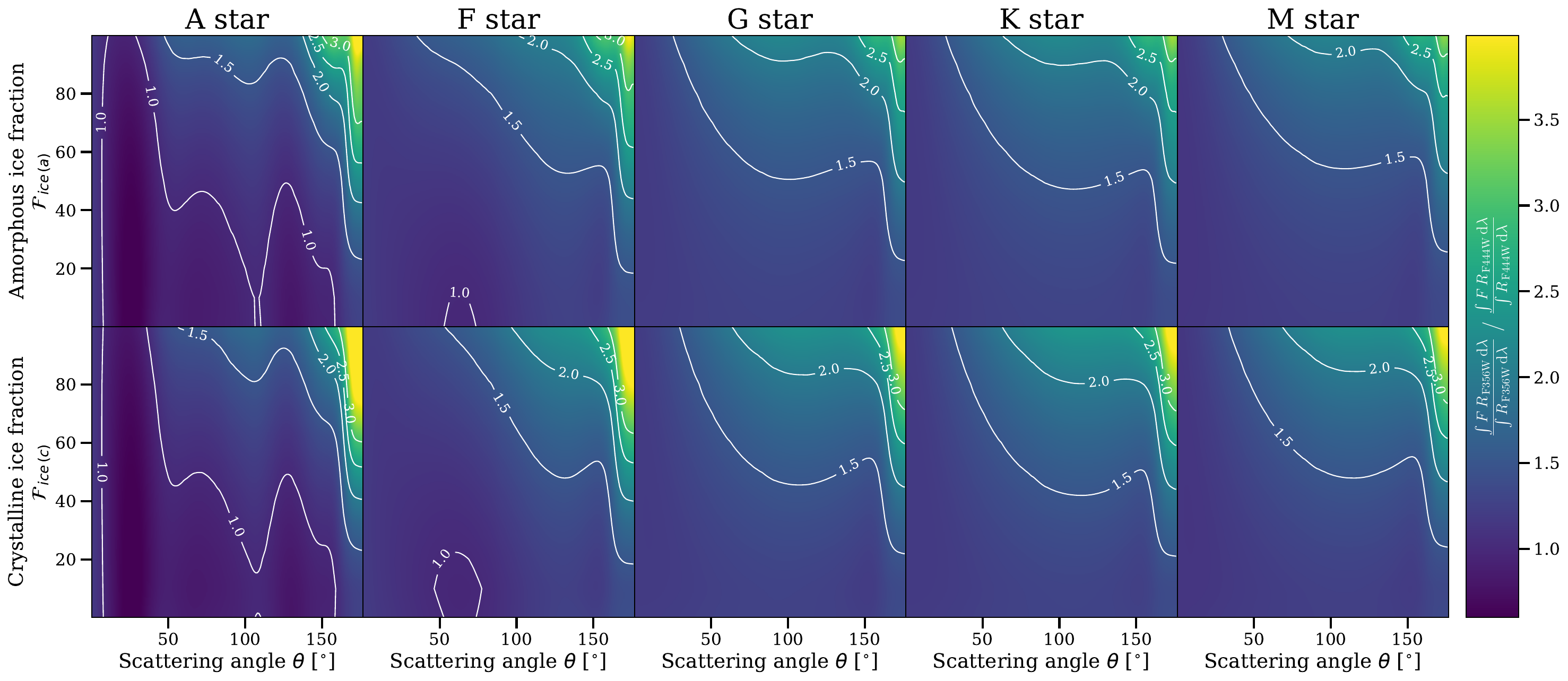}
\caption{The F356W/F444W ratio between SPFs of debris discs harbouring amorphous ice-dust aggregate (top panels) and crystalline ice-dust aggregate (bottom panels), as a function of the ice fraction ${\mathcal{F}}_{\rm ice}$ and spectral type depending on the scattering angle. See Sect.~\ref{sec: results-Phase functions} for details.}
\label{fig: SPF_photometry_grids_356_444}
\end{figure*}

\begin{figure*}
\includegraphics[width=18cm, height= 8cm]{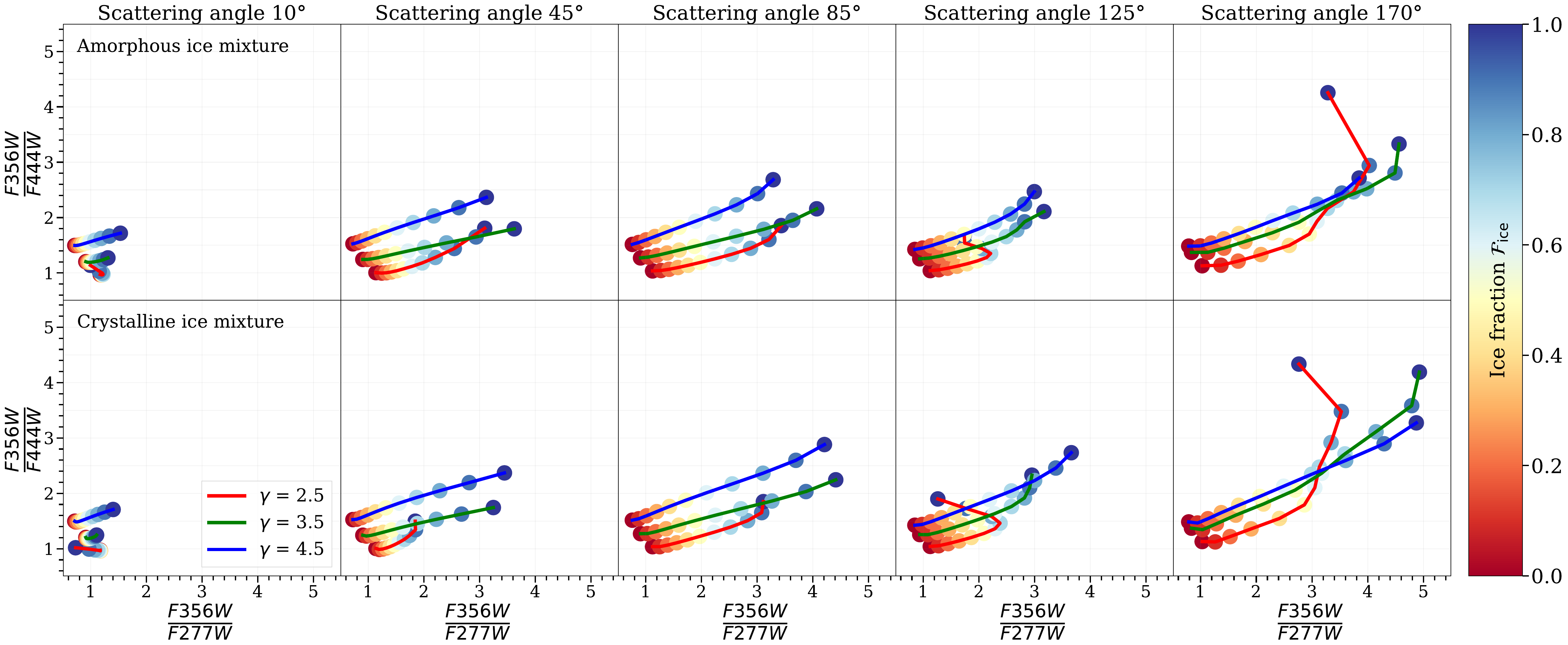}
\caption{Colour-colour diagrams comparing two filter pairs, F356W/F277W (y-axis) and F356W/F444W (x-axis), for debris discs around G-type stars. The plots show both amorphous (upper panel) and crystalline ice (lower panel) fractions as functions of scattering angle, varying with the power-law index $\gamma$ of the grain size distribution (indicated by red, green and blue lines) and ice fractions ${\mathcal{F}}_{\rm ice}$ (indicated by different colour dots with a colour bar). See Sections~\ref{sec: ice detection with JWST} and \ref{sec: methods} for details.}
\label{fig: filter_ice_gamma_sc}
\end{figure*}

Figs.~\ref{fig: SPF_photometry_grids_356_277} and \ref{fig: SPF_photometry_grids_356_444} present the simulated grid plots illustrating the ratio between the two most advantageous JWST photometric filter combinations (i.e., F277W and F356W as well as F356W and F444W) as a function of the ice fraction $\mathcal{F}_{\rm ice}$ and the central star's spectral type. We find that the ratio between these filter combinations exhibits remarkable consistency for debris discs around less luminous stars, such as G, K, and M types, particularly for low ice fractions (see also Fig.~\ref{fig: filter_ice_fraction_AGM_sc}), while A stars tend to have somewhat lower flux ratios. Furthermore, we find that photometric data alone do not provide clear constraints on the ice phase. Thus, spectroscopic observations are likely necessary to gain information about the ice phase.
 

As shown in Fig.~\ref{fig: SED_different_r_G_2}, the relative flux ratio varies not only with the grain size distribution (e.g., $\gamma$ values) but also with the ice fraction ${\mathcal{F}}_{\rm ice}$. To address the potential parameter degeneracy between ice abundance and size distribution (see also Sect.~\ref{sec: methods}), we present colour-colour diagrams in Fig.~\ref{fig: filter_ice_gamma_sc} to compare the two best filter pairs, F356W/F277W and F356W/F444W, for debris discs around G-type stars, varying with the power-law index $\gamma$ of the grain size distribution and ice fractions ${\mathcal{F}}_{\rm ice}$ for both amorphous (upper panel) and crystalline ice (lower panel) fractions as functions of scattering angle.

We first find that both F356W/F277W and F356W/F444W filter ratios are affected by both the $\gamma$ values and the ice fraction ${\mathcal{F}}_{\rm ice}$. The F356W/F277W ratio shows a stronger dependence on ice fraction ${\mathcal{F}}_{\rm ice}$ relative to its dependence on $\gamma$, with this effect being more pronounced for moderate to higher $\gamma$ values (e.g., $\gamma > 3.5$). In particular, the F356W/F444W ratio exhibits greater sensitivity to changes in $\gamma$ compared to the F356W/F277W ratio, which is more pronounced for lower $\gamma$ values (e.g., $\gamma$ = 2.5) and higher ice fractions ${\mathcal{F}}_{\rm ice}$. Consequently, these models suggest that the F356W and F444W filter combination is in fact optimal for constraining grain size distribution ($\gamma$ values), while the F356W and F277W filter combination is better for determining the ice fraction ${\mathcal{F}}_{\rm ice}$ in debris discs. These findings however suggest that two filters are not sufficient to strongly constrain the ice fraction in debris discs, though this issue might be circumvented if the size distribution were constrained via other means (perhaps other scattered light data, or model-based assumptions/assertions). In terms of optimal disk geometry, the results of Fig. \ref{fig: spf_ratio_grids_horizontal_SNR} hold, but there is somewhat more power to discern among different combinations of size distribution and ice fraction for near 90$^\circ$ scattering angles.

Furthermore, our findings indicate that photometry alone cannot effectively distinguish between crystalline and amorphous ice due to the similarities in results across different ice phases.  This underscores again the necessity of spectroscopic observations for accurately determining the ice phase.



Fig.~\ref{fig: filter_ice_fraction_AGM_sc_MG} presents the photometric ratios with F356W/F277W, and F356W/F444W of debris discs around A, G, and M stars for both amorphous and crystalline ice fractions as a function of different scattering angles, using EMT MG rule. Building upon previous findings (see Sect.~\ref{sec: results-SED}), indicating that the choice of dust mixing rule has a minimal effect on the SEDs, we demonstrate that this insensitivity extends to photometric ratios, which show again nearly identical results across different mixing rules (see also Fig.~\ref{fig: filter_ice_fraction_AGM_sc}). Consequently, we conclude that key parameters such as the depth and location of ice features, as well as the JWST photometric measurements, remain essentially almost unaltered by the specific dust mixing rule employed, further simplifying the interpretation of observational data.

\section{Summary}\label{sec: summary and outlooks}

To investigate the characterisation of different forms of ice (amorphous and crystalline) in debris discs, and to provide useful limits on the existence, properties, and spatial distribution of ice in these discs with current observations such as JWST, we have conducted a numerical feasibility study. We assumed a range of fractional ice compositions within icy dust mixtures, considering the diverse structures of debris discs, including their spatial distribution and inclination across various stellar spectral types (A, F, G, K, and M). We quantified and discussed the impact of the ice fraction ${\mathcal{F}}_{\rm ice}$ and phase on the observational characteristics of debris discs, including detailed assessments of the resulting spectral energy distributions and scattering phase functions. We particularly focused on the shorter wavelength regime around 3 microns, which harbours a wealth of diagnostic ice features and where scattered light by dust grains dominates - a regime of particular interest for JWST observations. Our key findings are:

\begin{itemize}

\item [--] The strength of the $\sim$ 2.7 and 3.3\,$\mu$m ice features increases with increasing ice fraction, and their difference in relative flux densities becomes more pronounced. These ice features show a dependence on the scattering angle and disc inclination angle, with backscattering yielding the most prominent ice features. This contrast serves as a valuable indicator of the ice composition and distribution.

\item [--] The 3.1\,$\mu$m ice feature (so-called `Fresnel peak') serves as a viable discriminant for the transitional phase such as crystallinity, implying the exposure to the temperature, simultaneously offering a temperature probe. For discs around less luminous stars, the 3.25\,$\mu$m crystalline ice feature is particularly discernible in backscattering regimes with higher ice fractions. Furthermore, the peak of the $\sim$3.4\,$\mu$m feature can further serve as a diagnostic tool, shifting slightly by crystallinity, whose feature is more pronounced in larger scattering angle regimes (e.g., less inclined discs) around luminous stars.

\item [--] The SPFs of dust particles within debris discs provide crucial insights into the ice fraction ${\mathcal{F}}_{\rm ice}$. The forward scattering part of the phase function is less dependent on particle characteristics, while the backscattering part shows more pronounced variations. For observability with an estimated SNR, strong forward and backscattering is advantageous for higher ice fractions, particularly, discs around more luminous stars. At intermediate scattering angles, the expected SNR variation is weaker. In general, given that observations of highly inclined discs are harder to obtain and interpret, disc inclination should not be a strong driver of target choice.

\item [--] We present simulated JWST photometry ratios that are relevant for detecting ice features in debris discs, depending on the ice fraction ${\mathcal{F}}_{\rm ice}$ and scattering angles to provide useful observational constraints for assessing the detectability of ice in debris discs. Since each filter incorporates trends associated with varying ice fractions, considering multiple filters can be advantageous. Any combination involving the F277W filter and other filters provides favourable constraints on the ice fraction $\mathcal{F}_{\rm ice}$, including the minimum detectable ice fraction. Particularly, the combination between the F277W and F356W NIRCam filters, as well as the combination between the F356W and F444W NIRCam filters, constitute advantageous filter combinations for constraining the ice fraction within the observed dust grains in debris discs. 


\item [--] Our models show that there is a degeneracy between the grain size distribution and ice fraction for photometric flux ratios. The F356W/F444W ratio shows greater sensitivity to changes in $\gamma$ compared to the F356W/F277W ratio, while the F356W/F277W ratio exhibits a stronger dependence on ice fraction ${\mathcal{F}}_{\rm ice}$ relative to its dependence on $\gamma$. Obtaining strong constraints on these parameters will require more than two filters, or if possible, spectra.


\item [--] The overall spectral features (e.g., location and strength of ice absorption features), along with the photometric data obtained by JWST, show negligible variation regardless of the dust mixing approximation applied.

\end{itemize}

\section*{Acknowledgements}
The authors are grateful to Prof. Sebastian Wolf for insightful discussions on modelling debris discs for the present study. MK gratefully acknowledges funding from the Royal Society. GMK is supported by the Royal Society as a Royal Society University Research Fellow. VR acknowledges the support of the Italian National Institute of Astrophysics (INAF) through the INAF GTO Grant ``ERIS \& SHARK GTO data exploitation''.
This work was partly supported by the European Union’s Horizon 2020 research and innovation program and the European Research Council via the ERC Synergy Grant ``ECOGAL'' (project ID 855130).

\section*{Data Availability}
The data underlying this study are available in the article or are available in the referenced papers.

\bibliographystyle{mnras}
\bibliography{ice_debris_disc} 

\appendix

\section{Optical properties (scattering and absorption cross sections ($C_{\rm sca}$ and $C_{\rm abs}$) of dust grains}\label{app: scattering and absorption cross sections}

\begin{figure*}
\includegraphics[width=17.5cm, height= 6.8cm]{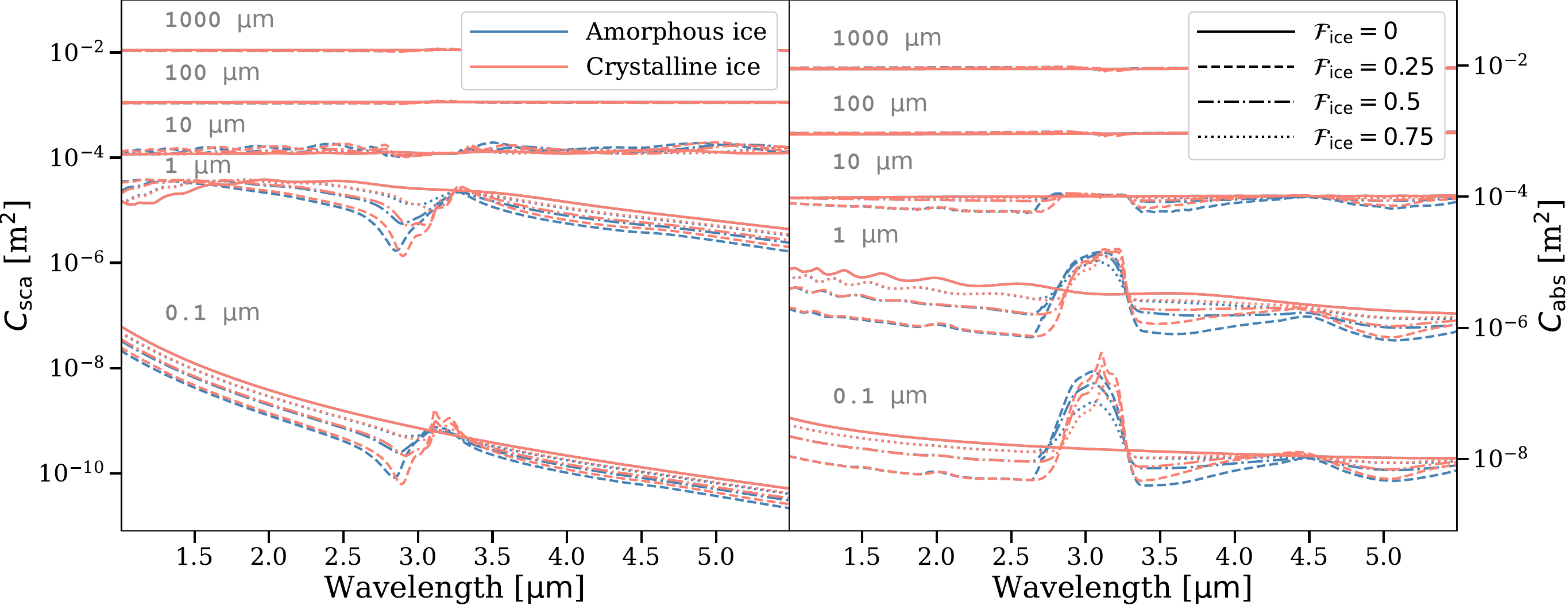}
\caption{Simulated scattering and absorption cross sections ($C_{\rm sca}$ and $C_{\rm abs}$, respectively) of amorphous ice (blue lines) and crystalline ice (red lines) for different grain sizes depending on the volume fraction of ice ${\mathcal{F}}_{\rm ice}$, using rules of EMT Brugemann  (\citealp{Bruggeman1935}). See Fig.~\ref{fig: n_k} and Sect.~\ref{sec: model} for details.}
\label{fig: cabs_sca_sil_ice}
\end{figure*}

\begin{figure*}
\includegraphics[width=17.5cm, height= 6.8cm]{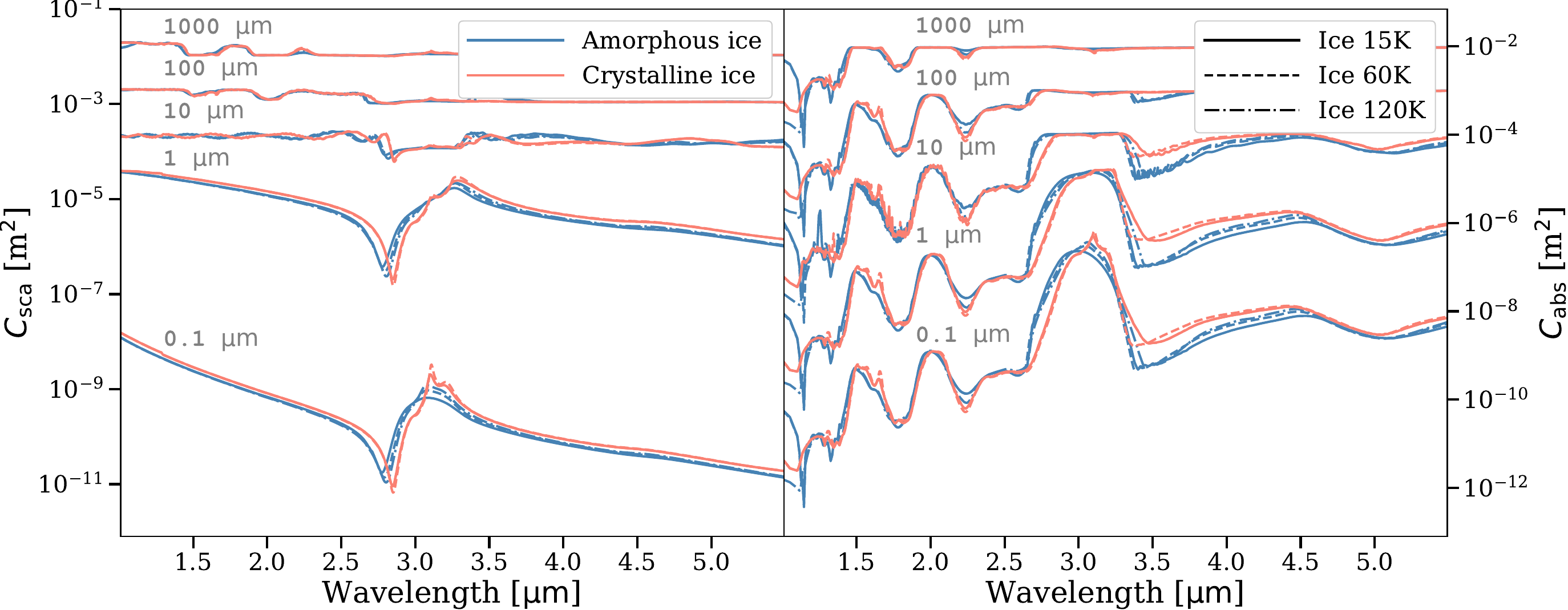}
\caption{Simulated scattering and absorption cross sections ($C_{\rm sca}$ and $C_{\rm abs}$, respectively) of amorphous ice (blue lines) and crystalline ice (red lines) for different grain sizes depending on the temperature of ice ${\mathcal{F}}_{\rm ice}$. See Fig.~\ref{fig: n_k} for details.}
\label{fig: cabs_sca_ice}
\end{figure*}

Characteristic features as a function of ice phase and temperature (see Fig.~\ref{fig: n_k}) are reflected in the scattering and absorption cross-sections $C_{\rm {\,sca}}$ and $C_{\rm {\,abs}}$ (see Figs.~\ref{fig: cabs_sca_sil_ice} and \ref{fig: cabs_sca_ice}), which are distinctly represented in the corresponding features of the resulting SEDs. See Sect.~\ref{sec: results-SED} for details.

\section{Photometry with JWST filters}\label{app: Photometry with JWST filters}

\begin{figure*}
\includegraphics[width=18cm, height= 18cm]{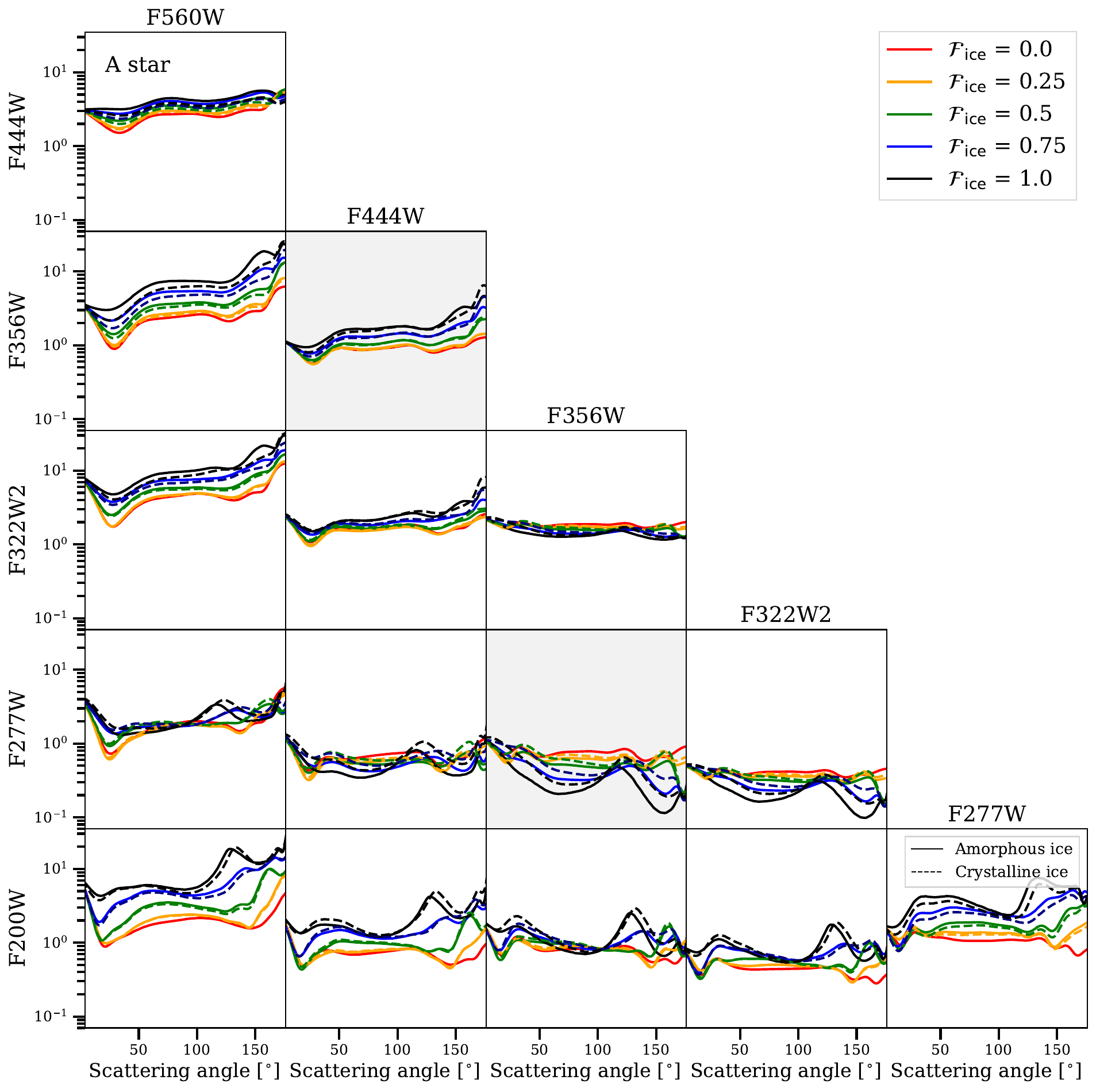}
\caption{Calculated corner plots of JWST photometric measurements of debris discs around A star through a set of JWST filters for different ice fractions $\mathcal{F}_{\rm ice}$ and phases. The filters for calculating the ratio are labelled on the Y-axis for the denominator and the X-axis for the numerator, respectively. Shaded regions, e.g., the combination between F356W and F444W, F277W and F356W, show a higher sensitivity to variations in the ice fraction across the entire range of scattering angles considered in our simulations, indicating the advantageous filter combinations. See Sect.~\ref{sec: ice detection with JWST} for details.}
\label{fig: corner_plot_A_sc}
\end{figure*}

\begin{figure*}
\includegraphics[width=18cm, height= 18cm]{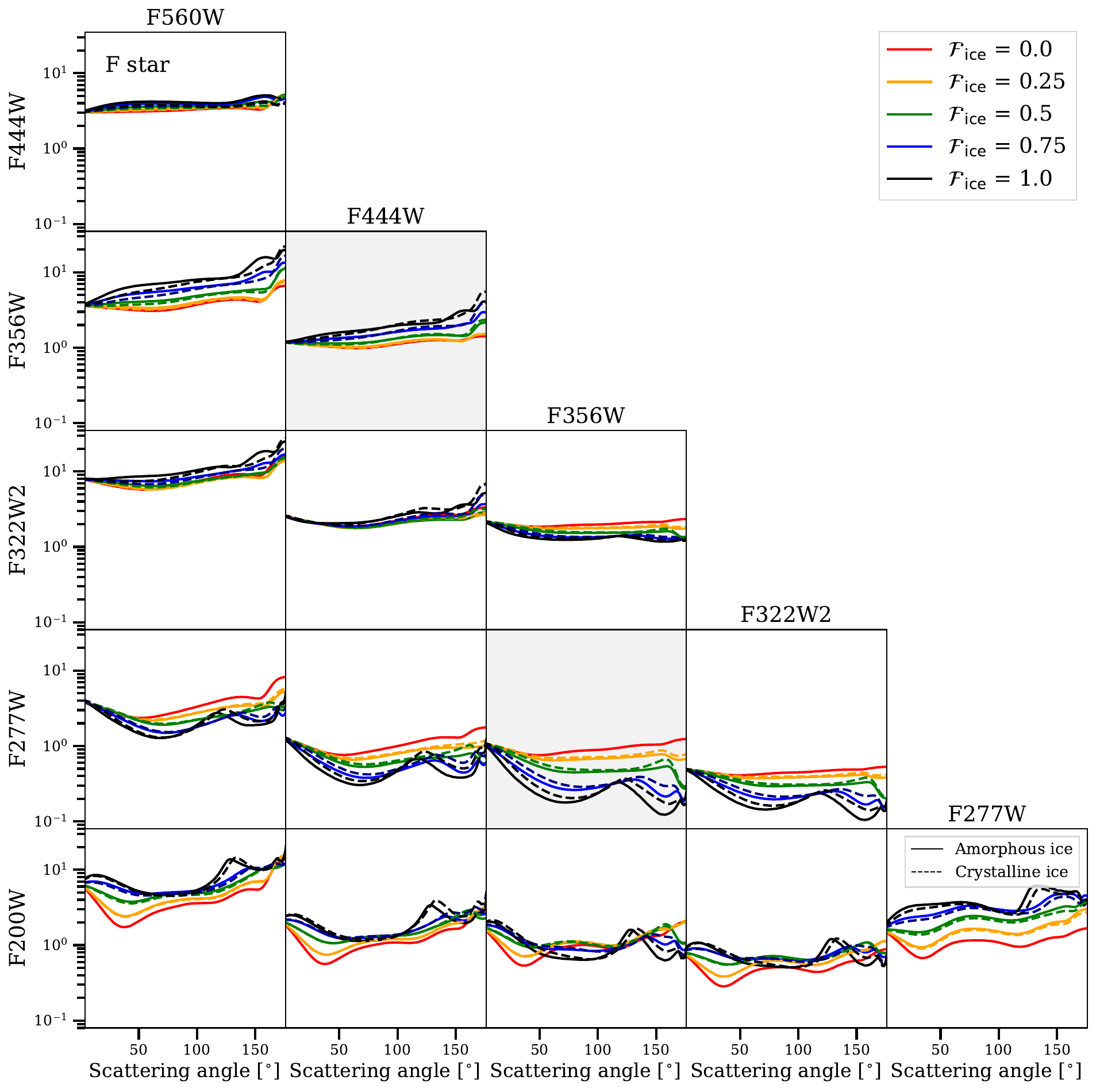}
\caption{Calculated corner plots of JWST photometric measurements of debris discs around F star through a set of JWST filters for different ice fractions $\mathcal{F}_{\rm ice}$ and phases. The filters for calculating the ratio are labelled on the Y-axis for the denominator and the X-axis for the numerator, respectively. Shaded regions, e.g., the combination between F356W and F444W, F277W and F356W, show a higher sensitivity to variations in the ice fraction across the entire range of scattering angles considered in our simulations, indicating the advantageous filter combinations. See Sect.~\ref{sec: ice detection with JWST} for details.}
\label{fig: corner_plot_F_sc}
\end{figure*}

\begin{figure*}
\includegraphics[width=18cm, height= 18cm]{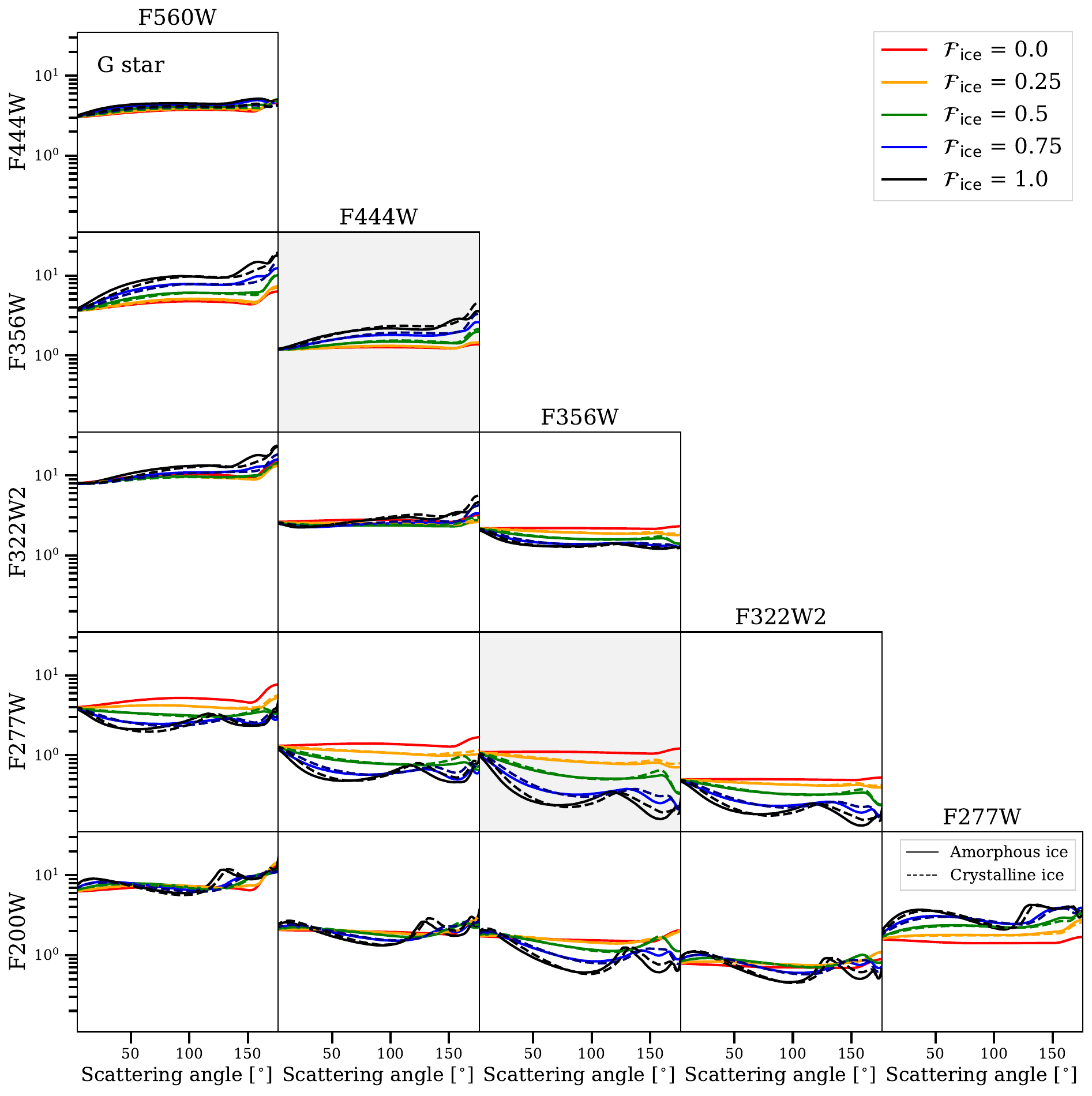}
\caption{Calculated corner plots of JWST photometric measurements of debris discs around G star through a set of JWST filters for different ice fractions $\mathcal{F}_{\rm ice}$ and phases. The filters for calculating the ratio are labelled on the Y-axis for the denominator and the X-axis for the numerator, respectively. Shaded regions, e.g., the combination between F356W and F444W, F277W and F356W, show a higher sensitivity to variations in the ice fraction across the entire range of scattering angles considered in our simulations, indicating the advantageous filter combinations. See Sect.~\ref{sec: ice detection with JWST} for details.}
\label{fig: corner_plot_G_sc}
\end{figure*}

\begin{figure*}
\includegraphics[width=18cm, height= 18cm]{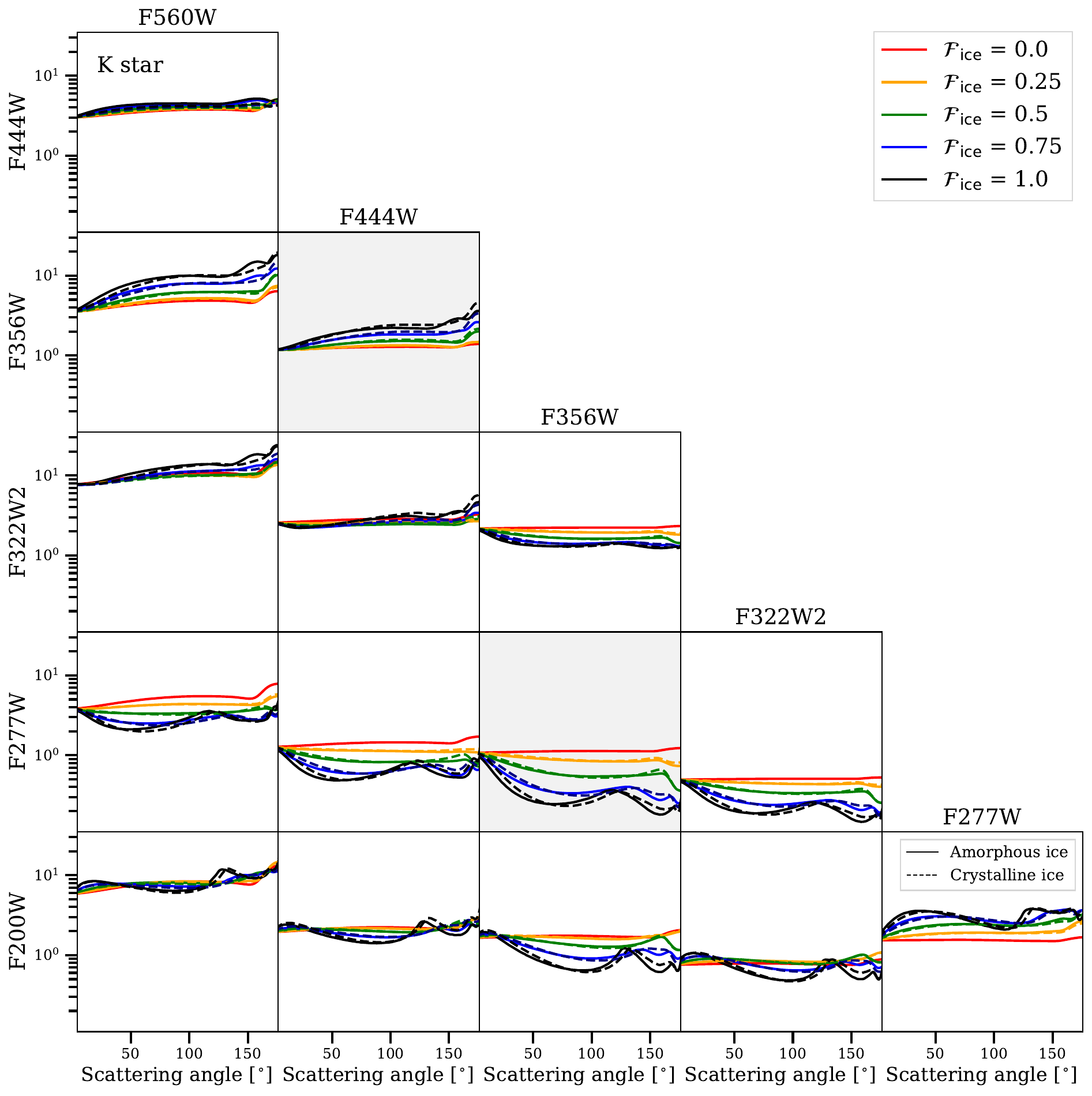}
\caption{Calculated corner plots of JWST photometric measurements of debris discs around K star through a set of JWST filters for different ice fractions $\mathcal{F}_{\rm ice}$ and phases. The filters for calculating the ratio are labelled on the Y-axis for the denominator and the X-axis for the numerator, respectively. Shaded regions, e.g., the combination between F356W and F444W, F277W and F356W, show a higher sensitivity to variations in the ice fraction across the entire range of scattering angles considered in our simulations, indicating the advantageous filter combinations. See Sect.~\ref{sec: ice detection with JWST} for details.}
\label{fig: corner_plot_K_sc}
\end{figure*}

\begin{figure*}
\includegraphics[width=18cm, height= 18cm]{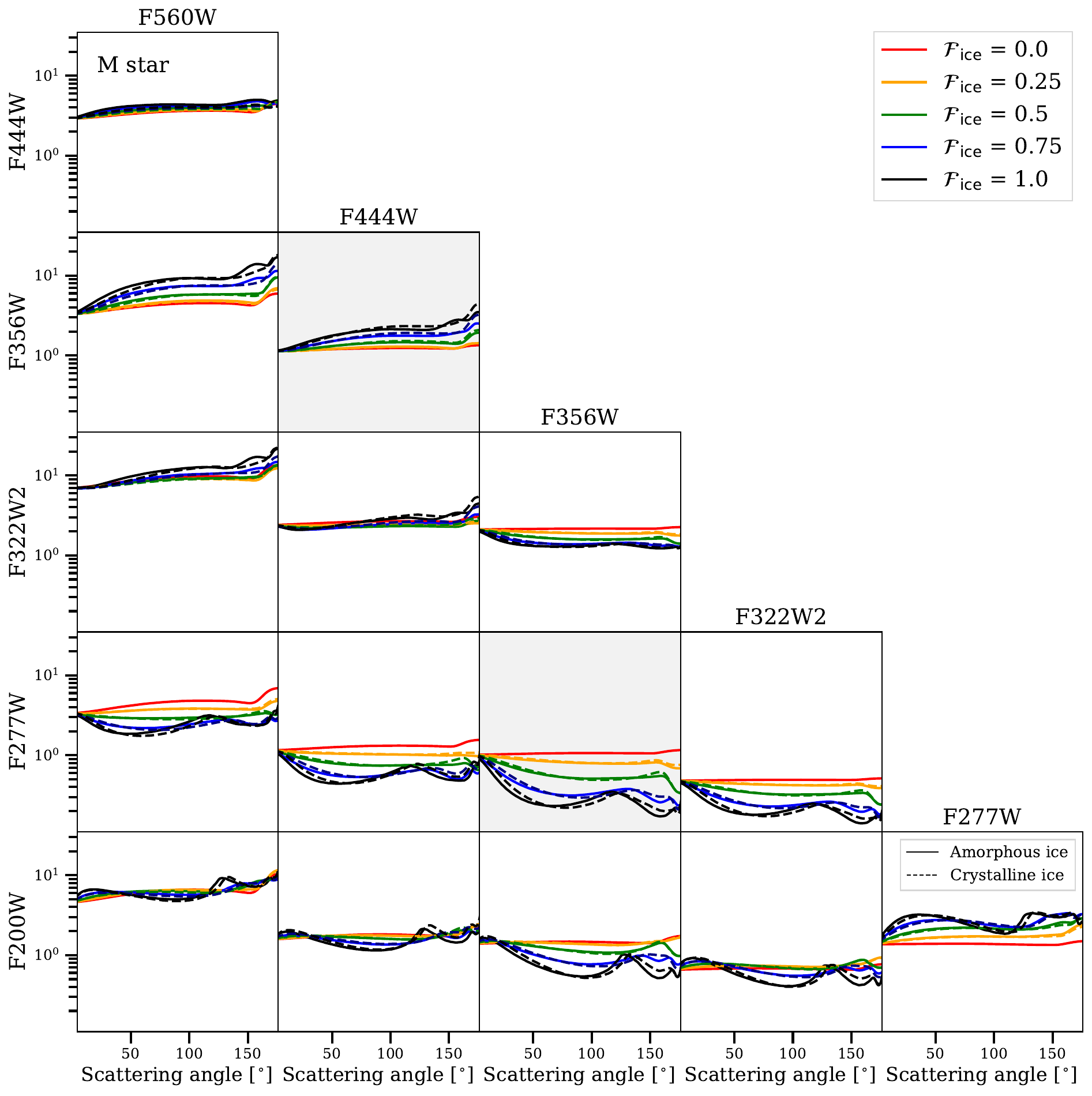}
\caption{Calculated corner plots of JWST photometric measurements of debris discs around M star through a set of JWST filters for different ice fractions $\mathcal{F}_{\rm ice}$ and phases. The filters for calculating the ratio are labelled on the Y-axis for the denominator and the X-axis for the numerator, respectively. Shaded regions, e.g., the combination between F356W and F444W, F277W and F356W, show a higher sensitivity to variations in the ice fraction across the entire range of scattering angles considered in our simulations, indicating the advantageous filter combinations. See Sect.~\ref{sec: ice detection with JWST} for details.}
\label{fig: corner_plot_M_sc}
\end{figure*}

\section{Selection of mixing rules}\label{app: Selection of mixing rules}

\begin{figure*} 
\includegraphics[width=17.5cm, height= 22cm]{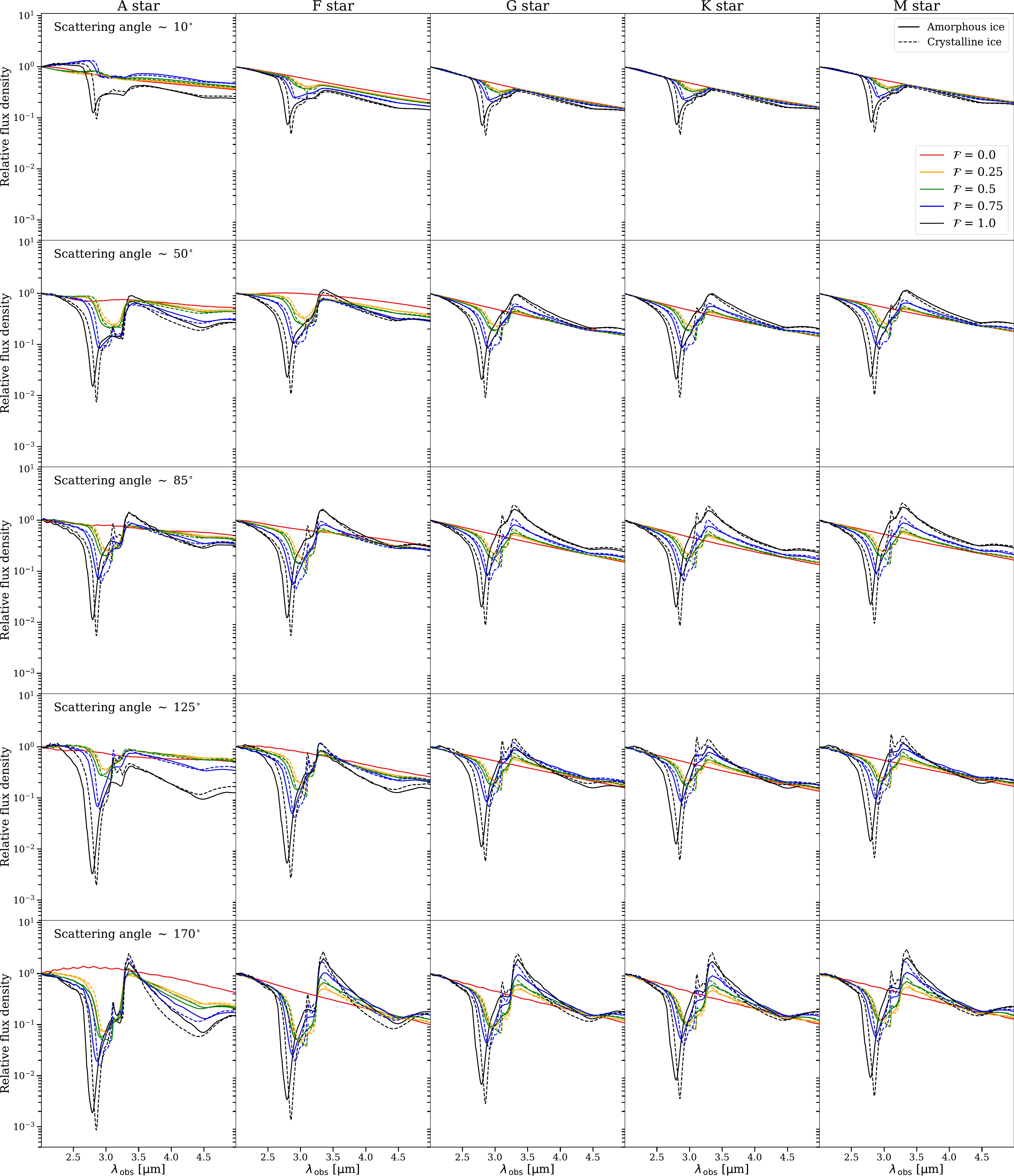}
\caption{Relative flux density of SEDs of exterior debris discs at 50 au as a function of the volume fraction of ice ${\mathcal{F}}_{\rm ice}$ and spectral type of the central star depending on scattering angles, using Brueggemann mixing rule (\citealp{Bruggeman1935}).  All spectra are normalised to 1 at 2.0\,$\mu$m. See Sect.~\ref{sec: results-SED} for details.}
\label{fig: SED_AFGKM_40au_RF_sa_MG}
\end{figure*}

\begin{figure*}
\includegraphics[width = 16cm, height = 10cm]{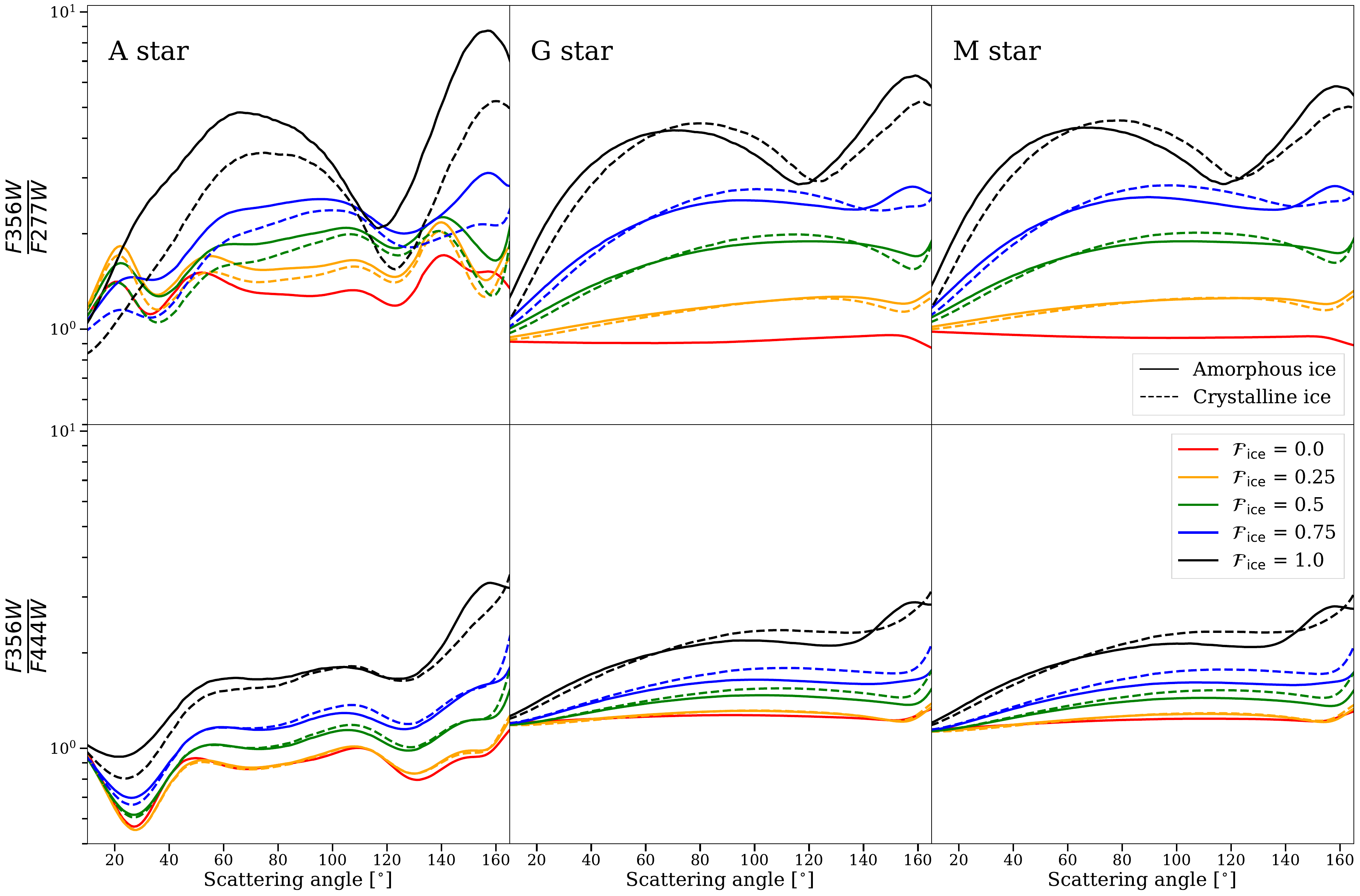}
\caption{Ratio between F356W and F277W (upper panel) and F356W and F444W (lower panel) of debris discs around A, G, and M stars for both amorphous and crystalline ice fractions as a function of different scattering angles, using EMT Maxwell-Garnett rule (\citealp{Maxwell-Garnett1906}). See Sects.~\ref{sec: methods} and \ref{sec: ice detection with JWST} for details.}
\label{fig: filter_ice_fraction_AGM_sc_MG}
\end{figure*}

\begin{figure*}
\includegraphics[width = 16cm, height = 10cm]{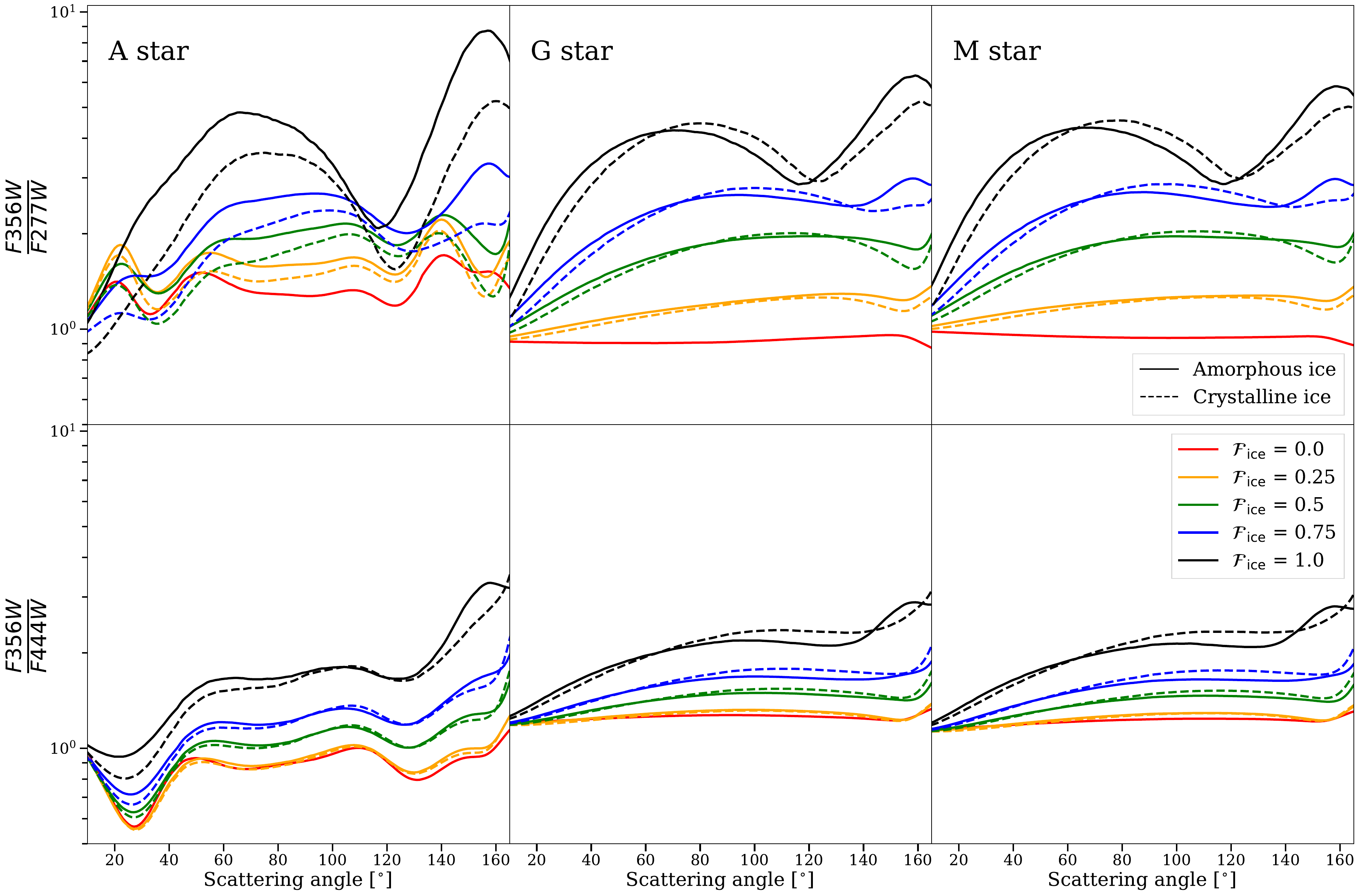}
\caption{Ratio between F356W and F277W (upper panel; this value is flipped compared to the one, i.e., F277W/F356W shown in Figs. \ref{fig: corner_plot_A_sc}, \ref{fig: corner_plot_F_sc}, \ref{fig: corner_plot_G_sc}, \ref{fig: corner_plot_K_sc}, and \ref{fig: corner_plot_M_sc}) and F356W and F444W (lower panel) of debris discs around A, G, and M stars for both amorphous and crystalline ice fractions as a function of different scattering angles, using Bruggeman rule (\citealp{Bruggeman1935}). See Sect.~\ref{sec: ice detection with JWST} and Appendix \ref{app: Photometry with JWST filters} for details.}
\label{fig: filter_ice_fraction_AGM_sc}
\end{figure*}

\label{lastpage}
\end{document}